\documentclass[conference]{IEEEtran}
\ifCLASSINFOpdf

\else
\fi

\let\labelindent\relax
\usepackage{nicefrac}
\usepackage{siunitx}
\usepackage{array,framed}
\usepackage{booktabs}
\usepackage{
  color,
  float,
  epsfig,
  wrapfig,
  graphics,
  graphicx,
  subcaption
}
\usepackage{setspace}
\usepackage{latexsym,fancyhdr,url}
\usepackage{enumerate}

\usepackage[linesnumbered,ruled, vlined]{algorithm2e}
\usepackage{algorithmic}
\usepackage{bbm}
\usepackage{graphics}
\usepackage{xparse} % argument parsing -- \edist
\usepackage{xspace}
\usepackage{csvsimple}
\usepackage{balance}
\usepackage{color}
\usepackage{tikz}
\usepackage{amsmath}
\usepackage{multirow}
\usepackage{listings}
\usepackage{graphicx}
\usepackage{comment}
\usepackage{pifont}
\usepackage{tabularx}
\usepackage{tabu}
\usepackage{array}
\usepackage{varwidth}
\usepackage{makecell}
\usepackage{enumitem}
\usepackage{threeparttable}
\usepackage{upgreek}
\usepackage{amsfonts}
\usepackage{amsmath,amsfonts,bm}

\def\eqref#1{equation~\ref{#1}}
\def\1{\bm{1}}

\def\vd{{\bm{d}}}

\def\vr{{\bm{r}}}
\def\vs{{\bm{s}}}

\def\vz{{\bm{z}}}

\def\mH{{\bm{H}}}

\DeclareMathAlphabet{\mathsfit}{\encodingdefault}{\sfdefault}{m}{sl}
\SetMathAlphabet{\mathsfit}{bold}{\encodingdefault}{\sfdefault}{bx}{n}

\newcommand{\stitle}[1]{\vspace{1ex}\noindent{\bf #1}}

\usepackage{
  tikz,
  pgfplots,
  pgfplotstable
}
\usepackage{hyperref}
\definecolor{darkblue}{rgb}{0.0, 0.0, 0.55}
\definecolor{darkcandyapplered}{rgb}{0.64, 0.0, 0.0}
\hypersetup{
    colorlinks=true,
    linkcolor=darkblue,
    citecolor=darkcandyapplered,
    filecolor=magenta,      
    urlcolor=darkblue,
}
\usepackage{dirtytalk}

\usetikzlibrary{
  shapes.geometric,
  arrows,
  external,
  pgfplots.groupplots,
  matrix
}

\definecolor{mygray}{gray}{0.9}
\definecolor{dkgreen}{rgb}{0,0.6,0}
\definecolor{mauve}{rgb}{0.58,0,0.82}
\newcommand{\Code}[1]{\lstinline{#1}}

\newcommand{\ie}{\mbox{\emph{i.e.,\ }}}
\newcommand{\eg}{\mbox{\emph{e.g.,\ }}}

\pgfplotsset{compat=1.9}
\newcommand{\sys}{{\tt CLExtract}\xspace}

\newcommand{\nip}[1]{\vspace{1ex}\noindent\textbf{#1}}

\usepackage{mathtools}

\DeclareMathAlphabet{\mathcal}{OMS}{cmsy}{m}{n}
\DeclareGraphicsExtensions{%
    .png,.PNG,%
    .pdf,.PDF,%
    .jpg,.mps,.jpeg,.jbig2,.jb2,.JPG,.JPEG,.JBIG2,.JB2}

\begin{document}

\title{CLExtract: Recovering Highly Corrupted DVB/GSE Satellite Stream with Contrastive Learning }

\author{Anonymous Author(s)}

\author{\IEEEauthorblockN{Minghao Lin}
\IEEEauthorblockA{University of Colorado Boulder\\
minghao.lin@colorado.edu}
\and
\IEEEauthorblockN{Minghao Cheng}
\IEEEauthorblockA{Independent Researcher\\
minhalemail@gmail.com}
\and
\IEEEauthorblockN{Dongsheng Luo}
\IEEEauthorblockA{Florida International University\\
dluo@fiu.edu}
\and
\IEEEauthorblockN{Yueqi Chen}
\IEEEauthorblockA{University of Colorado Boulder\\
yueqi.chen@colorado.edu}}

% \IEEEoverridecommandlockouts
% \makeatletter\def\@IEEEpubidpullup{6.5\baselineskip}\makeatother
% \IEEEpubid{\parbox{\columnwidth}{
%     {\fontsize{7.5}{7.5}\selectfont Preprint}
% }
% \hspace{\columnsep}\makebox[\columnwidth]{}}

\maketitle

\maketitle
% Kernel is the most important software component of an operating system (OS), and the security community has been striving to proactively detect and patch kernel vulnerabilities. 

\begin{abstract}
Since satellite systems are playing an increasingly important role in our civilization, their security and privacy weaknesses are more and more concerned. 
For example, prior work demonstrates that the communication channel between maritime VSAT and ground segment can be eavesdropped on using consumer-grade equipment. 
The stream decoder GSExtract developed in this prior work performs well for most packets but shows incapacity for corrupted streams. 
We discovered that such stream corruption commonly exists in not only Europe and North Atlantic areas but also Asian areas. 
In our experiment, using GSExtract, we are only able to decode 2.1\% satellite streams we eavesdropped on in Asia.

Therefore, in this work, we propose to use a contrastive learning technique with data augmentation to decode and recover such highly corrupted streams.
Rather than rely on critical information in corrupted streams to search for headers and perform decoding, contrastive learning directly learns the features of packet headers at different protocol layers and identifies them in a stream sequence. 
By filtering them out, we can extract the innermost data payload for further analysis.
Our evaluation shows that this new approach can successfully recover 71-99\% eavesdropped data hundreds of times faster speed than GSExtract. 
Besides, the effectiveness of our approach is not largely damaged when stream corruption becomes more severe.

\end{abstract}

\section{Introduction}
\label{sec:intro}
Satellite systems are becoming infrastructures of modern civilization.
It provides a wide range of services including media broadcasts that cover 100 million customers, Earth observation which contributes to environmental conservation efforts, and precise global positioning services.
Especially in recent years, the New Space trend~\cite{newspace} significantly advances the development of organizations such as SpaceX and OneWeb that carry space missions like global broadband service.
Nowadays, there are more than 2,000 operational satellites that orbit Earth and the market value exceeds \$150 billion a year~\cite{market}.

A satellite system consists of three major components: ground segment, space segment, and ground-space communication. 
While all components are reported vulnerable in previous works (\eg \cite{icarus}\cite{debris}\cite{cybserspaceinnewspace}\cite{viasat}), for adversaries, ground-space communication is the most accessible attack surface.
Unlike firmware in ground-segment and satellite payload the reverse engineering of which requires physical access, ground-space communication relies on radio signals and covers a large area in the size of million square kilometers.
As long as there is an antenna within the area and aligned to the satellite, attackers can eavesdrop on satellite streams and steal sensitive data. 

A prior work~\cite{tale} demonstrates this threat by eavesdropping on maritime VSAT communications in the North Atlantic. 
The stream decoder GSExtract developed in this work can extract between 40-60\% of the GSE PDUs contained within the targeted streams.
But for corrupted packets, GSEtract shows incapacity by recovering only 10-25\% of them.
We discovered that such stream corruption commonly exists for satellite communication in not only European areas but also Asian areas.
% We further discovered that stream corruption is more severe for communication other than maritime VSAT.

In our experiment, using GSEtract, we are only able to decode 2.1\% satellite stream we eavesdropped on in Asia.
This higher corruption is presumably because of surface reflection which is not an issue for martime VSAT eavesdropping. 
However, in addition to surface reflection, we identified three more factors that universally influence satellite stream quality and are difficult to be eliminated.

The traditional Finite-State Machine (FSM) based decoding approach, like the one implemented in GSExtract, fundamentally cannot recover such highly-corrupted satellite streams.
This is because it relies on critical information in networking packet headers (\eg length field, CRC-32) to perform decoding layer by layer.
When this critical information is corrupted, the decoding can hardly carry on.

In this work, we propose to use a contrastive learning technique with data augmentation to recover satellite streams.
Instead of relying on critical information for decoding, contrastive learning directly learns the features of packet headers at different layers and identifies them in a stream sequence.
By filtering them out, we can extract the innermost payload that can be further analyzed by tools like Wireshark.
Our approach further employs data augmentation to entitle the trained contrastive learning model with robustness against unseen corruptions.
We implemented our approach and named it as \sys. 
Our experiment shows that \sys can successfully recover 71-99\% eavesdropped data hundreds of times faster speed than GSExtract.
Besides, the effectiveness of \sys is not largely damaged when corruption becomes more severe. 

Making eavesdropping more practical, we develop \sys to facilitate investigation in security and privacy issues in satellite communication.
To foster future research, we will open-source \sys once this work is accepted.
In summary, this paper makes the following contributions: 
\begin{itemize}
    \setlength\itemsep{0em}
	\item Analysis into factors that cause satellite stream corruption and challenges of corrupted stream recovery.
        \item Proposed contrastive learning technique with data augmentation to recover corrupted satellite streams.
	\item Open-sourced implementation of the proposed technique and evaluation of its effectiveness and efficiency using eavesdropped data in the real world
\end{itemize}

In the following, we first describe the background in Section~\ref{sec:bg}, followed by the design overview in Section~\ref{sec:overview}.
Then, we elaborate on the technical details in Section~\ref{sec:tech}. 
The evaluation results and ethics are in Section~\ref{sec:eval}.
Finally, we conclude our work in Section~\ref{sec:conclusion}.
 
%-------------------------------------------------------------------------------
\section{Background and Challenges}
\label{sec:bg}
In this section, we first introduce the protocol layers of the satellite stream and packet formats of each layer. 
Then, we describe four major factors that can corrupt the integrity of DVB/GSE packets, especially in the scenario of eavesdropping. 
Finally, we discuss the challenges of recovering corrupted streams.

\begin{figure}[t]
\centering
\includegraphics[width=\linewidth]{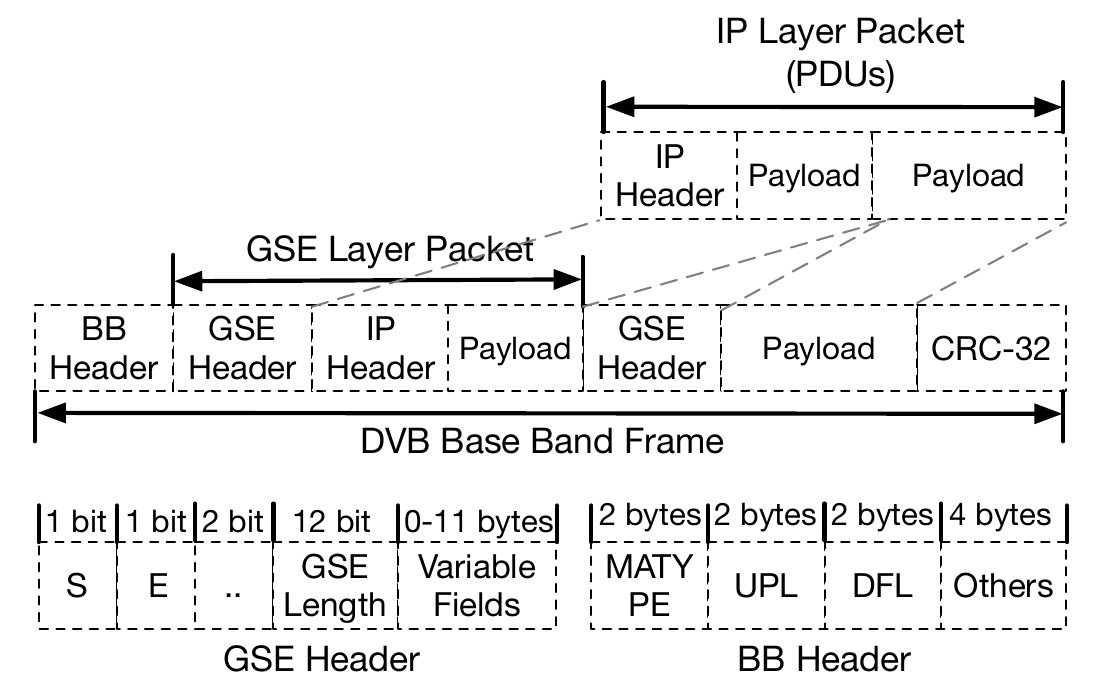}
\caption{Protocol layers of satellite stream and packet formats of each layer. }
\label{fig:protocol}
\vspace{-3ex}
\end{figure}

\nip{DVB/GSE Protocol Format.}
DVB (Digital Video Broadcasting) and its following DVB-S2 and S2X designed by ETSI (The European Telecommunications Standards Institute) are the de facto standard for the communication of most satellites. 
Data following this standard is encapsulated into continuous streams using GSE (Generic Stream Encapsulation) protocol. 
As shown in Figure~\ref{fig:protocol}, from innermost to outermost, the encapsulation consists of three protocol layers: IP layer, GSE layer, and DVB layer.

The IP layer supports common transport layer protocols (\eg TCP, ICMP, and MPEG). 
The IP header has a field recording which transport layer protocol is used (\eg 0x06 for TCP and 0x01 for ICMP). 
All transport layer data along with the IP header are encapsulated into PDUs (Protocol Data Units).
PDUs are further divided into several slots and stored as data fields in continuous GSE packets.
A GSE packet starts with a header of variable length. 
Figure~\ref{fig:protocol} shows its layout. 
The fixed header is in two bytes, recording whether the current GSE packet is the starting packet (\ie S=1) or the ending packet (\ie E=1) for a complete PDU. 
Besides, it includes a length field that stores the size of the current GSE packet.
Traditional FSM-based decoding approach relies on this length field to pinpoint the start position of the next GSE packet.
More information like Fragment ID and Protocol Type is stored in variable fields.
Due to the space limit, we omit details of this part.
Readers can refer to ~\cite{gse} for more information. 
In the ending GSE packet of a complete PDU, there is a CRC-32 tail as the error detection code of the PDU.

With all GSE packets encapsulated, at the outermost DVB layer, several GSE packets are concatenated to fill in the data field of a Base Band (BB) Frame.
The BB frame starts with a BB header the size of which are 10 bytes. 
The first two bytes of the header are the MATYPE field which describes the input stream format and the type of Mode Adaptation. 
The second two bytes are the UPL field storing user packet length in bits (up to 65535 bits). 
The following two bytes are  the DFL field which holds the length of the data field (\ie concatenated GSE packets) in bits (up to 58112 bits). 
Readers can refer to ~\cite{dvb} for details.

% \begin{figure}[t]
% \centering
% \includegraphics[width=.5\linewidth]{figs/factors.pdf}
% \caption{Four factors that cause packet corruption: ionosphere perturbation, extreme weather, surface reflection, and antenna quality. }
% \label{fig:factor}
% \vspace{-3ex}
% \end{figure}

\nip{Four Factors Causing Stream Corruption.}
Given a complete DVB/GSE packet, it is straightforward to decode it using a traditional FSM-based approach (\eg GSExtract\cite{tale}) which decodes layer by layer according to information in headers (\eg data field length).
The innermost PDUs (\ie IP packets) extracted in this approach can be further analyzed by tools like Wireshark.
However, the quality of radio communication between the satellite and the ground segment, especially in the scenario of eavesdropping, is under the influence of many factors, which makes the FSM-based approach can hardly carry on. 
% In Figure~\ref{fig:factor}, we illustrate four inevitable factors that cause packet corruption.

The first factor is solar activities which result in perturbations of the ionosphere. This perturbation can change the density structure of the ionosphere by creating areas of enhanced density. 
This change reflects, refracts, and absorbs radio waves, leading to the loss of signal. 
According to NASA's report~\cite{solar}, when communication is interrupted, some satellites can be tumbled out of control for several hours, and weather images can be lost. 
The second factor causing signal attenuation and absorption are rainstorms, snow, and heavy winds near the ground, given that the signals from satellites are transmitted through air~\cite{weather}. 
Higher-frequency bands tend to be more affected by rain because the wavelength itself is close in size to water molecules. 
The third factor is the surface reflection from distant reflectors such as mountains and large industrial infrastructures and local environmental effects including shadowing and blockage from objects and vegetation near the terminal~\cite{surface}. 
The last factor is the quality of the antenna. 
Attackers performing eavesdropping can seldom afford professional antennae and lack experience in antenna alignment.
Therefore, the DVB/GSE packets received through consumer-grade antenna are more likely to be of low quality in comparison with the ground station.

\nip{Challenges of Recovering Corrupted Stream.}
Due to the nature of satellite radio communication, the four factors are not easy to eliminate.
As we will show in the evaluation (Section~\ref{subsec:setup}), oftentimes, the DVB/GSE packets received by eavesdroppers are highly corrupted.
% The bit flipping and signal losing cause many troubles to packet decoding.
To make matters worse, eavesdroppers are passive attackers and cannot ask for re-transmission if discovering that the received packets are corrupted.
How to decode corrupted DVB/GSE packets and extract PDUs is the key to the success of eavesdropping.
In the following, we discuss technical challenges in detail.

The first challenge is broken stream. Eavesdropping may start from the  middle of a BB frame or because of bad weather, packet headers are missed. 
With such a broken stream, it is difficult to determine where to start decoding. 
One brute-force solution is to treat every byte as a potential header.
However, it is impractical in the real world because such decoding speed is too low to catch up with the transmission speed: with the best practice, we are only able to decode 609.71 MB in one hour, not identifying even one complete header in the highly corrupted stream we eavesdropped.

The second challenge is corrupted critical fields. 
One critical field is the length of packets.
Even if we are able to identify the BB header or GSE header, for decoding, we need to determine the length of data fields at each layer so as to extract PDUs. 
Unfortunately, these length fields are corrupted and thus untrustworthy. 
Besides, signal loss can happen. 
Therefore, we cannot use the size of inner packets to correct the length field in outer packets.
Without the right length, the decoding can hardly carry on.
Another critical field is the protocol type in IP packets because once it is corrupted, Wireshark cannot determine which transport layer protocol is used and fails to analyze PDUs.

The third challenge is nonfunctional CRC-32. The error detection code is designed to examine if PDUs are corrupted. 
In a clear and high-quality communication channel, CRC-32 helps find corruption.
However, our experiment results show that, in eavesdropping, corruption is so common that CRC-32 itself can also be problematic.
Therefore, we cannot rely on it to correct our decoding results.  
\section{Design Overview}
\label{sec:overview}

\begin{figure}[t]
    \centering
    \includegraphics[width=\linewidth]{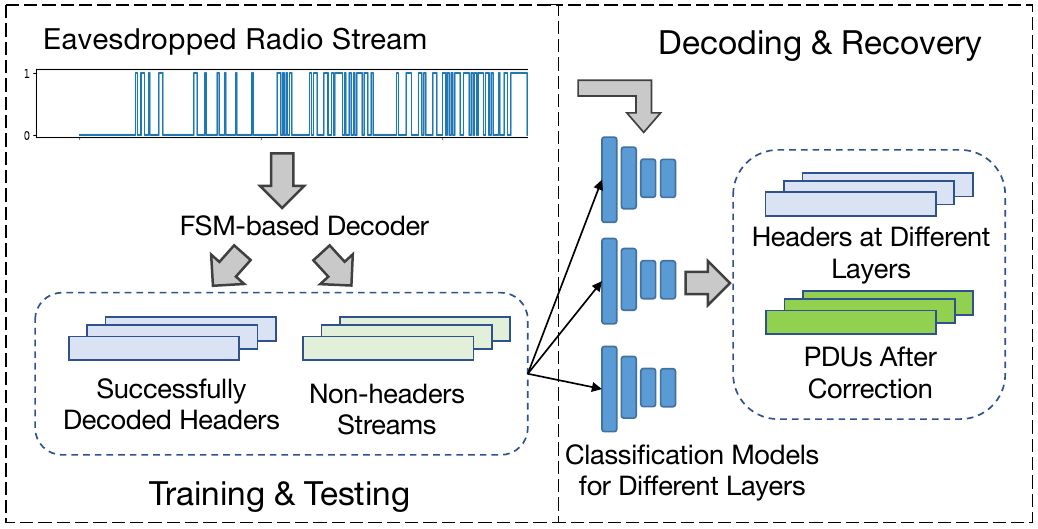}
    \caption{The recovering workflow. We first train a classification model using the eavesdropped data and then apply this model to identify headers even if they are corrupted and extract PDUs after correction.}
    \label{fig:pipeline}
        \vspace{-2ex}
\end{figure}

To resolve the technical challenges discussed above, in this work, we propose to employ state-of-the-art contrastive-learning techniques with data augmentations to decode and recover corrupted satellite streams.
Instead of relying on critical fields or CRC-32 for decoding, contrastive learning directly learns the features of packet headers at different layers and identifies them in a stream sequence.
By filtering them out, we can extract the innermost IP packets. 
Then we correct the transport layer protocol field in IP headers so that the extracted IP packets can be further analyzed by tools like Wireshark.
Figure~\ref{fig:pipeline} shows the workflow of our approach.

To start, we  first construct a training and testing dataset that consists of positive data (\ie successful decoded headers) and negative data (\ie non-headers streams).
Applying FSM-based GSExtract on the eavesdropped data, we collect positive data - BB headers, GSE headers, and IP headers that can be successfully decoded. 
Although successful decoding does not mean these headers are 100\% correct because the fields not used in FSM-based decoding can still be problematic, we deem these headers not corrupted and use them for training. 
This flaw in the training data, as we will describe in Section~\ref{sec:tech} and show in the evaluation (Section~\ref{sec:eval}), won't influence the accuracy of our approach with the assistance of a pre-trained encoder network. 
Eliminating headers from successfully decoded streams, we obtain negative data - non-headers streams. 
With these two types of streams, we train our classification model.
Given a sequence of bytes, this model can tell whether it is a header or not. 

We trained three instances of this classification model to identify BB headers, GSE headers, and IP headers, respectively. 
We first apply the BB header model to divide the whole eavesdropped streams into BB frames.
Within each BB frame, we apply the GSE header model to identify GSE layer packets.
Then, we apply the IP header model  to determine whether a GSE layer packet includes an IP header.
With the extracted IP packets in hand, we use Hamming distance to correct the transport layer protocol field in the IP header if it is corrupted.
After correction, the PDUs can be fed to tools like Wireshark for further analysis.

% \subsection{Header Extractor}
% Considering headers, including BB headers and GSE headers from the streaming data are highly corrupted, traditional finite-state machine (FSM) based methods, such as GSExtractor, may suffer from unbearable false negative instances. 
% To handle this problem, we propose to adopt state-of-the-art contrastive-learning methods with data augmentations, \sys, to detect headers from the eavesdropped data. 
% Figure~\ref{fig:pipeline} shows the training and detection workflow of the proposed \sys. 

\section{Technical Details}
\label{sec:tech}
In this section, we cover the technical details of our approach.
First, we present the framework of our contrastive learning based classification model.
Then, we take a scrutiny into the framework, elaborating on how it pre-trains an encoder network and how the encoder network is used to fine-tune a classification model.

\subsection{Contrastive Learning Based Classification Framework}
\begin{figure}[t]
    \centering
    \includegraphics[width=\linewidth]{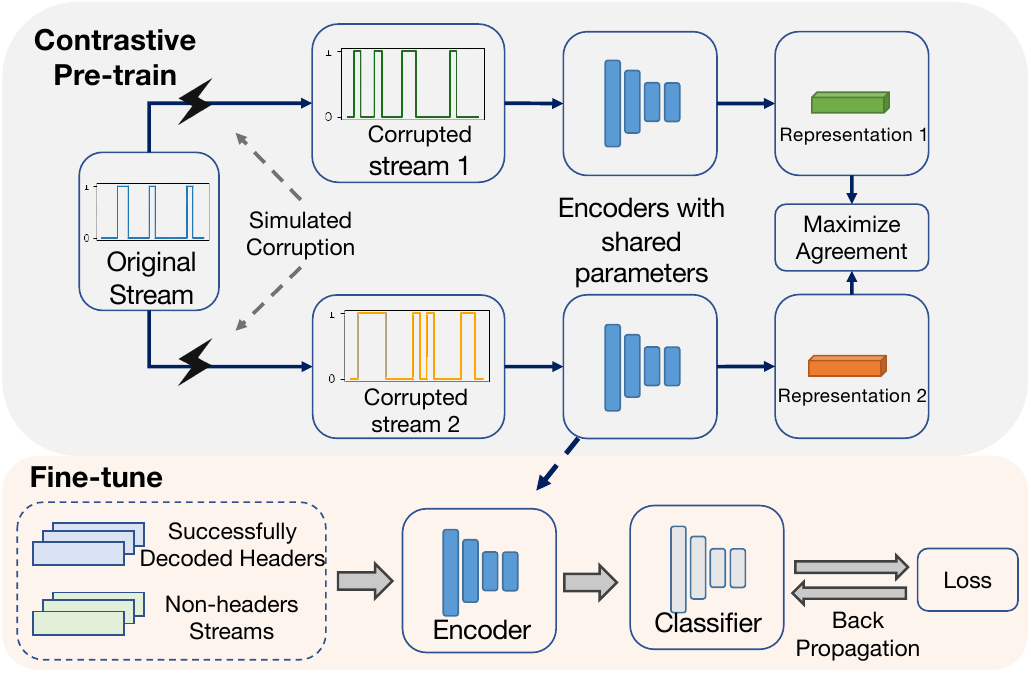}
    \caption{The framework of proposed contrastive learning based classification model. The framework first pre-trains an encoder network and then uses it to fine-tune a supervised classification model.}
    \label{fig:framework}
    \vspace{-2ex}
\end{figure}

As a state-of-the-art machine learning paradigm, contrastive learning has achieved great success in the computer vision~\cite{chen2020simple}\cite{he2020momentum} and natural language processing domains~\cite{yang2019xlnet}\cite{logeswaran2018efficient}. 
Figure~\ref{fig:framework} illustrates the overall framework of our proposed contrastive learning based classification model. 
The first component of the framework is a self-supervised contrastive pre-train which generates an encoder network.
This encoder network is used in the second component to fine-tune a classifier model which predicts whether the stream input is a header or not.

\begin{figure}[t]
    \centering
    \includegraphics[width=\linewidth]{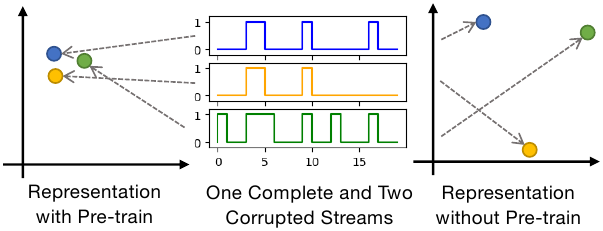}
    \caption{The diagram illustrating the necessity of pre-train for a robust encoder network.}
    \label{fig:encoder-illustrate}
        \vspace{-2ex}
\end{figure}

As the essential part of this framework, the encoder network learns a fixed dimensional vector representation for a stream input.
Without technical details, Figure~\ref{fig:encoder-illustrate} illustrates the difference between representation with and without a pre-trained encoder network.
A pre-trained encoder network can learn the features of non-corrupted headers in the meanwhile cluster corrupted headers with non-corrupted headers in the representation space.
In comparison, an encoder network without pre-train will disperse corrupted and non-corrupted headers in the representation space, mixing them with non-header streams and thus influencing the effectiveness of fine-tuned classification model.
Therefore, the encoder network with pre-train is robust to corruption noise, not only fixing the flaw of our training data mentioned in Section~\ref{sec:overview}, but also equipping the classification model with the capacity to identify corrupted headers.

% Self-supervised learning obtains supervision from the data itself via a semi-automatic process~\cite{liu2021self}. 
% With large-scale unlabeled datasets, self-supervised pret-raining approaches achieve competitive or even better performances when compared with supervised counterparts. 
% In a nutshell, contrastive learning methods generate novel and realistic-looking training views for input instances with data augmentation techniques. 
% Views of the same instance are considered positive pairs and the ones of different inputs form negative pairs. 
% The objective of contrastive learning encourages positive pairs to be concentrated and negative pairs to be diluted in the representation space. 
% With appropriate data augmentations simulating the corruption process, contrastive pretrain makes the model robust to noise and thus equips the model with the capacity to detect corrupted signals.

% Next, we s the encoder with multi-layer perceptions as the classifier to fine-tune the model with successfully decoded headers and randomly sampled non-header stream instances. 

\subsection{Encoder Network and Contrastive Pre-train}
\begin{figure}[t]
    \centering
    \includegraphics[width=0.48\textwidth]{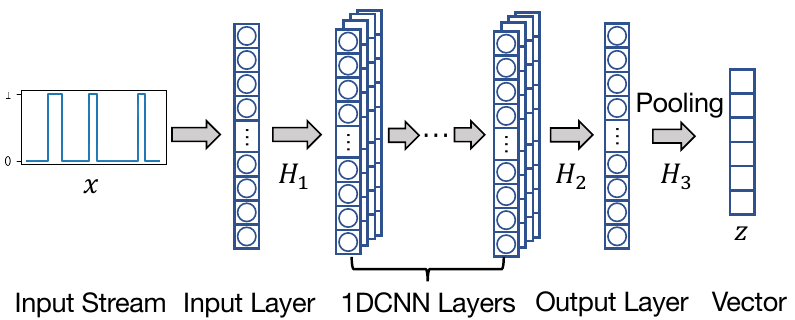}
    \caption{The encoder network that maps the radio stream to a fixed dimensional vector.}
    \label{fig:encoder}
    \vspace{-2ex}
\end{figure}

\stitle{Architecture of Encoder network.} 
The encoder network maps the stream data to a fixed dimensional vector. Formally, we denote it by $f_\theta(x): \mathbb{R}^{T}\rightarrow \mathbb{R}^{D}$. Here $x$ is the input stream; $T$ is the maximum length of possible headers; $D$ is the dimensionality of the representation vector; $\theta$ denotes the learnable parameters in the encoder network. Especially, for an input stream with a length smaller than $T$, we first pad it with 0x0 to make sure that all inputs are with the same length, \ie $T$. 
The architecture of the adopted encoder is shown in Figure~\ref{fig:encoder}.
It consists of a fully connected layer (input layer), a 10-layer dilated 1-dimensional convolutional neural network (1DCNN) module~\cite{yue2021learning}, a fully connected layer (output layer), and a pooling layer.
Compared to the vanilla 1DCNN, the dilated version has a large receptive field to capture long-range dependencies.  Formally, we have 
\begin{equation}
    \begin{aligned}
        \mH_1 = \text{InputLayer}(x)\\
        \mH_2 = \text{1DCNNs}(\mH_1)\\
        \mH_3 = \text{OutputLayer}(\mH_2)\\
        \vz = \text{MeanPooling}(\mH_3),
    \end{aligned}
\end{equation}
where $\mH_1,\mH_2,\mH_3$ are hidden representations, and $\vz\in \mathbb{R}^{D}$ is the representation vector.

\stitle{Data Augmentation by Simulating Corruption.}
To make the encoder network robust to corruption noise, we introduce data augmentation techniques. 
More specifically, we simulate corruption based on successfully decoded headers and add corrupted headers to pre-train dataset.
The communication corruption is usually modeled as white Gaussian noise.
The noise added to each bit is independent of the others. 
In general, there are two types of corruption. One is bit flip and another is bit loss.

We simulate flip and loss with parameterized ratios $\gamma_1, \gamma_2 \in [0,1]$. 
As shown in Algorithm~\ref{alg:flip}, we first randomly sample an array $\vd$ of $|x|$ real numbers from $(0,1)$ in Line 3, each corresponding to a bit in $x$. 
First, we determine if flip happens by comparing the sampled numbers with ratio $\gamma_1$. 
If $r_i<\gamma_1$, we flip the $i$-th bit in $x$ (line 4-6).
Then, we determine if loss happens by comparing $r_i$ with $\gamma_1$ and $\gamma_2$.
If $\gamma_1 < r_i < \gamma_2$,  we fill in the lost bit with 0 to simulate receivers (line 7-9). 
In other situations, we keep the bit. 
As such, the total corruption ratio is $\gamma_1+\gamma_2$ with $\gamma_1$ for flip and $\gamma_2$ for loss respectively.

% We compute the differences between $\vd$ and $\gamma$ in Line 4 and check the sign of the differences in line 5.  
% The final output $x'$ is obtained by considering both corrupted and uncorrupted bits (line 6). 
% The signal interruption is simulated with bit loss operation with a parameter $\gamma_2 \in [0,1) $ indicating the loss ratio. The algorithm is shown in Alg.~\ref{alg:bitloss}.

\begin{algorithm}[t]
    \centering
    \caption{Simulate Bit Flip \& Bit Loss}\label{alg:flip}
\begin{small}
    \begin{algorithmic}[1]
        \STATE {\bfseries Input:} a stream instance $x$, flip ratio $\gamma_1$, loss ratio $\gamma_2$;
        \STATE {\bfseries output:} Simulated corrupted stream instance $x'$;
        \STATE  $\vr \leftarrow$ random sample $|x|$ numbers from $\text{Uniform}((0,1))$;
        \STATE  $\vd_1 \leftarrow \vr- \gamma_1$;
        \STATE  $\vs_1 \leftarrow \text{sign}(\vd_1)$;
        \STATE  $x' \leftarrow  x*\vs_1 + (1-x)*(1-\vs_1)$ ; 
        \STATE  $\vd_2 \leftarrow \vr-(\gamma_1+\gamma_2)$;
        \STATE  $\vs_2 \leftarrow \text{sign}(\vd_2)$;
        \STATE  $x'' \leftarrow  x'*(1+\vs_2-\vs1)$ ; 
        % \STATE  $x'' \leftarrow  x'*\vs_2$ ; 
        \STATE  {\bfseries return } $x''$.
    \end{algorithmic}
\end{small}
\end{algorithm}

\stitle{Contrastive Pre-train.} 
Drawing inspiration from recent self-supervised learning algorithms in computer vision and natural language processing, we adopt, a simple but effective contrastive learning framework, SimCLR framework~\cite{chen2020simple} to learn representation vectors for input streams. 
SimCLR maximizes agreement between differently augmented streams of the same data example by using a contrastive loss in the latent space. Specifically, it contrasts the augmented streams generated from the same input (\ie positive pairs) by pulling them close in the representation space, while pushing apart the augmented streams generated from different inputs (\ie negative pairs).

Technically, for the contrastive pretrain objective, we follow SimCLR~\cite{chen2020simple} to use the normalized temperature-scaled cross entropy loss (NT-XEnt) as the contrastive loss. 
Algorithm~\ref{alg:pretrain} summarizes our contrastive pretrain procedure.
Specifically, we utilize the cosine similarity to define the similarity between two vectors $\vz_i$ and $\vz_j$. Formally,
\begin{equation}
    \text{s}(\vz_i,\vz_j) = \frac{\vz_i\cdot \vz_j}{||\vz_i||_2\cdot ||\vz_j||_2}
\end{equation}

Given a batch of $N$ stream instances, denote by $\{x_i\}_{i=1}^N$, we corrupt each instance twice to get $2N$ augmented streams. The ones corrupted from the same $x$ are considered positive pairs. For a positive pair $(\vz_i,\vz_j)$, we have its contrastive loss
\begin{equation}
    \ell_{(i,j)} = -\log\frac{\exp(\text{s}(\vz_i,\vz_j)/\tau)}{\sum_{k=1}^{2N} \mathbbm{1}_{[k\neq i]}\exp(\text{s}(\vz_i,\vz_k))/\tau)},
\end{equation}
where $\tau$ is the temperature parameter and $\mathbbm{1}_{[k\neq i]}$ is the indicator function defined as follows.
\begin{equation}
    \mathbbm{1}_{[k\neq i]} =  
    \begin{cases}
    0       & \quad \text{if } k=i\\
    1  & \quad \text{if } k \neq i.
  \end{cases}
\end{equation} 
Intuitively, minimizing $\ell_{(i,j)}$ encourages the model to identify the positive partner $\vz_j$ from $2N$ vectors for a given $\vz_i$. The batch loss is then computed by averaging all positive pairs in a mini-batch. Formally, we have 
\begin{equation}
\label{eq:batchloss}
   \mathcal{L} = \frac{1}{2N} \sum_{i=1}^N [\ell_{(2i-1,2i)}+\ell_{(2i,2i-1)}].
\end{equation}

\begin{algorithm}[t]
    \centering
    \caption{Contrastive Pre-train}\label{alg:pretrain}
\begin{small}
    \begin{algorithmic}[1]
        \STATE {\bfseries Input:} a set of stream instances $\{x_i\}$, batch size $N$, temperature $\tau$ ,encoder networks $f$;
        \STATE {\bfseries output:} Pretrained encoder networks $f$;
        \FOR{each minibatch $\{x_i\}_{i=1}^N$ of stream instances}
            \FOR{each instance $x_i$}
                \STATE $\tilde{x}_{i1} \leftarrow $ apply simulated corruptions on $x_i$;
                \STATE $\tilde{x}_{i2} \leftarrow $ apply simulated corruptions on $x_i$;
                \STATE $\vz_{2i-1} \leftarrow f(\tilde{x}_{i1})$;
                \STATE $\vz_{2i} \leftarrow f(\tilde{x}_{i2})$;
            \ENDFOR
            \STATE compute batch loss $\mathcal{L}$ with Eq.~(\ref{eq:batchloss});
            \STATE update parameters in $f$ by minimizing $\mathcal{L}$.
        \ENDFOR
    \STATE  {\bfseries return } encoder network $f$.
    \end{algorithmic}
\end{small}
\end{algorithm}
\vspace{-2ex}

\begin{algorithm}[t]
    \centering
    \caption{Supervised Fine-tune}\label{alg:finetune}
\begin{small}
    \begin{algorithmic}[1]
        \STATE {\bfseries Input:} a set of labeled stream instances $\{(x_i,y_i)\}$, batch size $N$, encoder networks $f$;
        \STATE {\bfseries Output:} classifier $g$;
        \STATE Inputlayer, 1DCNNs $ \leftarrow $ extract from encoder network $f$;
        \STATE classifier $\leftarrow$ initiate 2 fully connected layers
        \STATE detector $g \leftarrow$ stack Inputlayer, 1DCNNS, and classifier
        \FOR{each minibatch $\{(x_i,y_i)\}_{i=1}^N$}
            \FOR{each instance $x_i$}
                \STATE $\hat{y}_i \leftarrow g(x_i)$
                \STATE compute classification loss $\ell_i = \text{CrossEntropy}(\hat{y}_i,y_i))$;
            \ENDFOR
            \STATE compute batch loss $\mathcal{L}_{cls} = \frac{1}{N} \sum_{i=1}^N \ell_i$;
            \STATE update parameters in classifier by minimizing $\mathcal{L}_{cls}$.
        \ENDFOR
    \STATE  {\bfseries return } detector $g$.
    \end{algorithmic}
\end{small}
\end{algorithm}
\vspace{-2ex}

\subsection{Supervised Fine-tune and Classification Model}

After pre-train, the encoder network is robust to noise. We build our classifier on top of that to identify headers in the corrupted streams.
For efficiency, we extract the Inputlayer and 1DCNNs from the pre-trained encoder network and stack them with 2 fully connected layers as the classifier. 
Then, we fine-tune the classification model with supervised successfully decoded headers, labeled with 1, and randomly sampled non-header streams, labeled with 0. 
During this step, we kept the Inputlayer and 1DCNNs frozen and don't update parameters in these layers. 
Since we have two labels, headers and non-headers, we adopt the binary cross entropy as the loss function. 
Finally, we fine-tune the model under supervision using Algorithm~\ref{alg:finetune}. 

To recover the corrupted transport layer protocol field in IP headers, we adopt Hamming distance to calculate the distance between the potentially corrupted protocol value and the non-corrupted protocol value in the training dataset. The protocol value is automatically corrected to the target protocol with the smallest distance.

\section{Evaluation and Ethics}
\label{sec:eval}

\subsection{Experiment Setup and Ethics}
\label{subsec:setup}

In our experiment, we built a platform to eavesdrop on satellite communication data.
This platform consists of a TBS-6903 Professional DVB-S2 Dual Tuner PCIe Card and a professionally customized Ku-band antenna that can automatically align the dish. 
The total cost of this platform is around \$15k. 
We deployed the platform in a suburban area of a metropolis with more than 10 million of population in Asia, eavesdropping on the spectrum range from 11 GHZ to 12.75 GHZ which covers seven commercial satellites.
The eavesdropping was conducted over 20 days from July to October.
Finally, we received 23.6 GB DVB/GSE stream data. 
% We implement our approach on the basis of \fixme{xxx} and name it after \sys.

Considering that sensitive information could be included in the stream, we followed ethical principles proposed by the prior work~\cite{tale}, not storing any data longer than necessary. 
Even for the learning model trained using the eavesdropped data, we deleted it immediately after completing the evaluation to prevent adversary data generation from the model (\eg GAN~\cite{gan}). 
In the experiment, we treated data units in IP packets as normal payloads and didn't make any attempt to decrypt them.
For the sake of anonymity, we chose not to reveal the specific names of satellites and service providers that were eavesdropped on. 
Though we will open-source our implementation \sys after this work is accepted, we withhold the publishing of training data. 
This is our best effort to bolster future research.
Here, we advocate service providers or authorities to build a NVD-like databse (National Vulnerability Database)~\cite{nvd} which stores anonymized and desensitized data.
It will significantly advance research on space security.

From the eavesdropped data, we ran GSExtract to extract BB frames that can be successfully decoded in an FSM-based approach.
From these BB frames, we filter out BB headers, GSE headers, and IP headers - successfully decoded headers. 
Eliminating these headers, we obtain non-header streams.
The two types of data construct our training and testing dataset.
Using this approach, we collected in total 10471 BB frames (0.5 GB) from all streams (23.6 GB).
This low success decoding rate (2.1\%) indicates that corruption is very common in satellite communication. 
From such highly corrupted streams, as we will further show in our evaluation, tools like GSExtract which are built upon the traditional FSM-based decoding approach can only recover a very small portion of data.

% The training data is divided into two parts. The first part is the Non corrupted Headers. 
% Through the GSExtractor, we extract the data stream eavesdropped and get the complete bbframes. 
% Then we take out the corresponding GseHeaders and BBheaders from these bbframes. 
% In addition, by observing the structure diagram of the bbframe, we find that the first GseHeader is always located after the BBheader. 
% Therefore, in order to improve the accuracy of our model analysis, we also extract the combination of BBheader and the first GseHeader as supplementary training data. 

% The second part is the Non-headers. On the one hand, we use the non-header part of the complete bbframe extracted by the GSExtractor to take out the data of the length of GseHeader and BBheader as the training set.
% Here, since the length of the GseHeader is variable, ranging from 2 to 13 bytes, here we calculate the probability of the length of the GseHeader appearing in the complete bbframes, and corresponding to the calculated probability, the data of the corresponding length is taken out as Non-gseheader. 
% \dongsheng{Each experiment was conducted in 10 rounds. Average results are reported.}
With the training and testing data in hand, we evaluate \sys.
We divided the whole data set into two parts: 2/3 of successfully decoded headers and non-headers stream to train our model, and the remaining 1/3 to measure its \textbf{effectiveness}.
During data augmentation in contrastive pre-train, we set the flip ratio and loss ratio as 10\% and 10\% respectively.
To \textbf{compare} \sys with GSExtract, we apply both to streams with different corruption degrees and corruption types.
More specifically, we synthesize corruption using the same algorithm for data augmentation.
Starting from 2\%, we gradually increase the corruption degree $(\gamma_1+\gamma_2)$  by a step of 2\%, until 20\%.
In the meanwhile, we adjust the relative ratio flip($\gamma_1$) : loss($\gamma_2$) to 1:3, 1:1, and 3:1 to examine whether the types of corruption affect the robustness.
Since the model is trained over successfully decoded streams with data augmentation, there is no label leakage to apply the model for corrupted streams which can be considered as a distinct dataset. 
To evaluate to which extent \textbf{pre-trained network} with \textbf{data augmentation} improves the robustness, we compare the effectiveness of \sys with and without data augmentation.
Finally, we run \sys and GSExtract using a server with an Intel(R) Xeon(R) Gold 6258R CPU @ 2.70GHz, 1.5TB RAM, and an NVIDIA A100 80GB PCIe GPU, showing the efficiency of \sys. 
Each experiment mentioned above is repeated for 10 rounds and we report the average results.

% On the other hand, we simulate the bit flipping and loss that may occur in the satellite data transmission in the real world, starting from 0\%, increasing the probability of the possible corrupt bit in the complete bbframes' GseHeaders and BBheader by two percentage points. 

% If the bit is selected to be corrupted, then the probability of 1/4, 1/2 and 3/4 is set to select whether the bit is flipped or lost, and the Non-headers are obtained.

% \fixme{Do we mention the size of IP packets?}

\subsection{Evaluation Results of \sys}
\label{subsec:eval-result}

\nip{Effectiveness.}
We use four metrics to measure the effectiveness of \sys. The first metric is ACC which stands for the ratio of the number of correct predictions to the total number of input samples. In our scenario, it is calculated as \Code{(TP+TN)/(TP+TN+FP+FN)}. 
True positive (\Code{TP}) means the headers identified by \sys are indeed headers, and true negative (\Code{TN}) means \sys doesn't mistakenly identify non-header streams as headers. The second metric is Precision which is calculated as \Code{TP/(TP+FP)}. The third metric is Recall which is \Code{TP/(TP+TN)}. The last metric is F1 - the harmonic mean of Precision (\Code{P}) and Recall (R). It is calculated as (\Code{2/F1 = 1/P + 1/R}). F1 is high if and only if both Precision and Recall are high.

\begin{table*}[t]
    \centering
  \scalebox{0.8}{
    \begin{tabular}{c|c|c|c|c|c|c|c|c|c|c}
    \toprule
         Corruption   & \multirow{2}{*}{0.02} & \multirow{2}{*}{0.04} & \multirow{2}{*}{0.06} & \multirow{2}{*}{0.08} & \multirow{2}{*}{0.10} & \multirow{2}{*}{0.12} & \multirow{2}{*}{0.14} & \multirow{2}{*}{0.16} & \multirow{2}{*}{0.18} & \multirow{2}{*}{0.20}  \\
        Degree ($\gamma_1+\gamma_2$) & &  & &  &  &  &  &  &  & \\
        \midrule
         & \multicolumn{9}{c}{BB Headers}\\
        ACC & 0.984 & 0.981 & 0.976 & 0.971 & 0.964 & 0.955 & 0.946 & 0.935 & 0.921 & 0.914\\
        Precision & 0.977 & 0.976 & 0.974 & 0.973 & 0.973 & 0.969 & 0.966 & 0.962 & 0.959 & 0.957\\
        Recall   & 0.992 & 0.986 & 0.979 & 0.969 & 0.957 & 0.941 & 0.927 & 0.908 & 0.883 & 0.870 \\
        F1   &0.985 & 0.981 & 0.977 & 0.971 & 0.965 & 0.955 & 0.946 & 0.934 & 0.920 & 0.912\\
    \midrule
        & \multicolumn{9}{c}{GSE Headers}\\
             ACC & 0.883 & 0.870 & 0.854 & 0.839 & 0.828 & 0.813 & 0.797 & 0.784 & 0.769 & 0.757\\
        Precision &0.851 & 0.845 & 0.838 & 0.831 & 0.828 & 0.820 & 0.812 & 0.804 & 0.792 & 0.788  \\
        Recall  &0.933 & 0.910 & 0.882 & 0.856 & 0.833 & 0.807 & 0.780 & 0.758 & 0.737 & 0.713\\
        F1 & 0.890 & 0.877 & 0.859 & 0.843 & 0.830 & 0.813 & 0.796 & 0.780 & 0.764 & 0.748\\
    \midrule
        & \multicolumn{9}{c}{IP Headers}\\
             ACC & 0.970 & 0.971 & 0.972 & 0.971 & 0.972 & 0.971 & 0.971 & 0.971 & 0.970 & 0.969 \\
        Precision &0.962 & 0.962 & 0.962 & 0.961 & 0.962 & 0.960 & 0.960 & 0.960 & 0.957 & 0.954\\
        Recall  &0.980 & 0.981 & 0.983 & 0.982 & 0.982 & 0.983 & 0.984 & 0.984 & 0.985 & 0.985\\
        F1 &   0.971 & 0.972 & 0.972 & 0.972 & 0.972 & 0.971 & 0.972 & 0.972 & 0.971 & 0.969\\
       \bottomrule
    \end{tabular}
    }
    
    \parbox{0.68\textwidth}{
    \vspace{1ex}
    \caption{Effectiveness and robustness of \sys with different corruption degrees ($\gamma_1 + \gamma_2$) while the ratio of corruption types is fixed to $\gamma_1:\gamma_2 = 1:1$.}
    \label{tab:effectiveness}
    }
    \vspace{-4ex}
\end{table*}

In Table~\ref{tab:effectiveness}, we show the effectiveness of \sys when the relative corruption ratio flip($\gamma_1$) : loss($\gamma_2$)  is 1:1.
Overall, the four metrics for header identification at different layers are all very high no matter how corrupted the stream is - all above 0.71.3 while most are over 0.9. This indicates that \sys can accurately identify headers in a corrupted satellite stream.

From the table, we can further observe that, in comparison with BB headers and IP headers, the effectiveness of \sys in identifying GSE headers is relatively poor.
The reasons are two-fold.
On the one hand, unlike BB headers, the length of GSE headers is variable, ranging from 2 bytes to 12 bytes. 
In training, we have to pad the short GSE headers with 0x0 so that they can be uniformly represented in  the encoder network. 
However, the padded 0x0s carry no information and mislead \sys in classification if the non-header stream contains too many 0x0s.
On the other hand, unlike IP headers the length of which is at least 20 bytes, the length of GSE headers is relatively small.
Therefore, the classification model fails to learn enough features in training. 

Enough if the effectiveness of \sys in identifying GSE headers is not as good as identifying BB headers and IP headers, in comparison with GSExtract, \sys is much better.
In Table~\ref{tab:comparison}, we compare the numbers of identified GSE headers using \sys with the number using GSExtract. 
Due to the variable length and the small size nature mentioned above, the number of identified GSE headers by \sys is slightly lower than GSExtract when the corruption degree is 0.02. 
However, when corruption becomes more severe, regardless of the ratio of corruption types, \sys can successfully identify much more GSE headers than GSExtract (15707 vs. 516 when $\gamma_1+\gamma_2$=20\% and $\gamma_1:\gamma_2$=3:1 ).

\begin{figure}[t]
    \centering
    \begin{subfigure}[b]{0.26\textwidth}
         \includegraphics[width=1.4in]{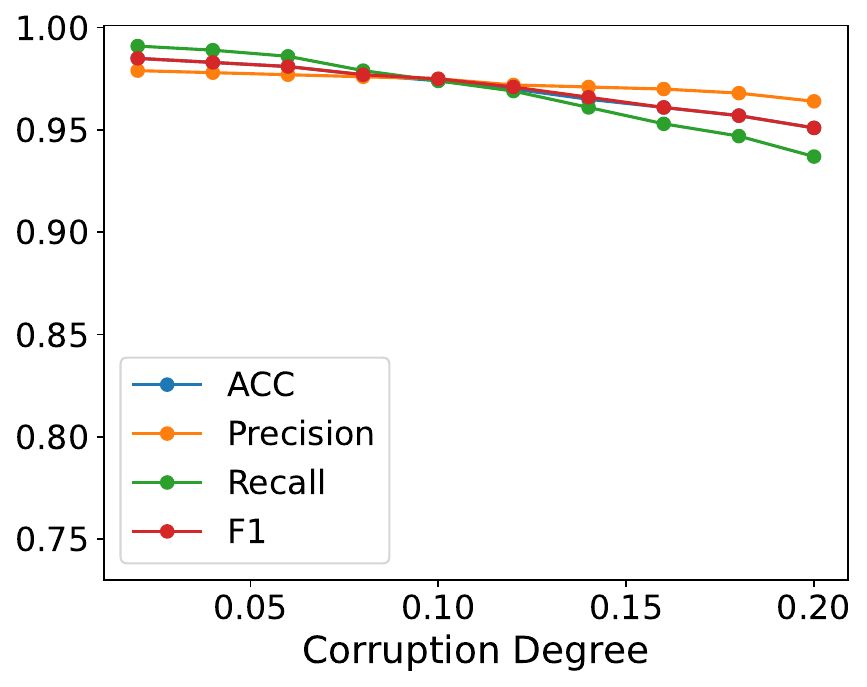}
         \caption{$\gamma_1$:$\gamma_2$ = $1:3$}
         \label{fig:exp1:bb:r25}
     \end{subfigure}
    \hspace{-3em}
    \begin{subfigure}[b]{0.26\textwidth}
         \includegraphics[width=1.4in]{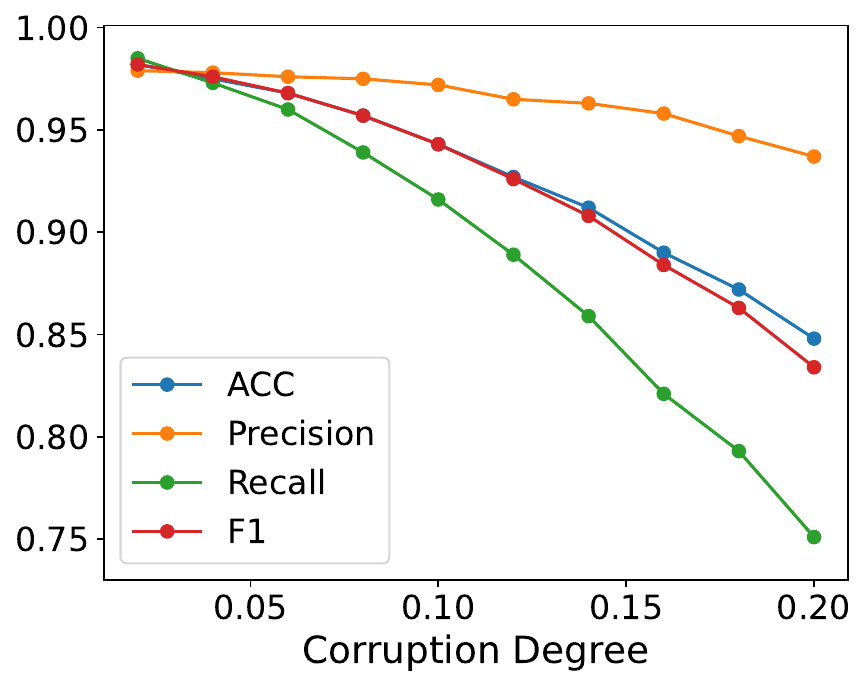}
         \caption{$\gamma_1$:$\gamma_2$ = $3:1$}
         \label{fig:exp1:bb:r75}
     \end{subfigure}
    \parbox{0.48\textwidth}{
    \caption{Corresponds to Table~\ref{tab:effectiveness}, effectiveness and robustness of BB header identification with different corruption degrees ($\gamma_1 + \gamma_2$) and ratios ($\gamma_1 : \gamma_2$). More corresponding results are in Figure~\ref{fig:exp1:gse} and \ref{fig:exp1:ip}. }
    \label{fig:exp1:bb}
    }
    \vspace{-3ex}
\end{figure}

As Table~\ref{tab:effectiveness} shows the results when the ratio of corruption types flip($\gamma_1$) : loss($\gamma_2$) is 1:1, we present the results for different ratios in Figure~\ref{fig:exp1:bb} for BB headers, Figure~\ref{fig:exp1:gse} for GSE headers, and Figure~\ref{fig:exp1:ip} for IP Headers (Figure~\ref{fig:exp1:gse} and \ref{fig:exp1:ip} are in Appendix~\ref{appendix}). 
From the three figures, we can observe the effectiveness of \sys in recovering corrupted streams, which aligns with the results in Table~\ref{tab:effectiveness} and ~\ref{tab:comparison}.

\begin{table*}[t]
    \centering
  \scalebox{0.8}{
    \begin{tabular}{c|c|c|c|c|c|c|c|c|c|c}
    \toprule
         Corruption   & \multirow{2}{*}{0.02} & \multirow{2}{*}{0.04} & \multirow{2}{*}{0.06} & \multirow{2}{*}{0.08} & \multirow{2}{*}{0.10} & \multirow{2}{*}{0.12} & \multirow{2}{*}{0.14} & \multirow{2}{*}{0.16} & \multirow{2}{*}{0.18} & \multirow{2}{*}{0.20}  \\
         Degree ($\gamma_1+\gamma_2$) & &  & &  &  &  &  &  &  & \\
    %     \midrule
    %     & \multicolumn{9}{c}{\# of BB / GSE / IP Headers with $\gamma_1:\gamma_2 = 1:1$} \\
    %     Ground-truth & &  & &  &  &  &  &  &  & \\
    %     GSExtract &7147/17905/7150 &5067/13639/5067 &3531/9849/3534 &2590/7231/2593  &1894/5070/1896  &1476/3838/1478  &1067/2738/1068  &840/1997/840  &667/1535/667  &506/1114/506 \\
    %     CLExtract &9999/16684/8588 &9831/16663/8587  &9670/16637/8587  &9521/16601/8588 &9264/16541/8588  &9116/16444/8589  &8857/16461/8588  &8587/16340/8588  &8394/16253/8588  &8081/16275/8588 \\
    % \midrule
    %     & \multicolumn{9}{c}{\# of BB / GSE / IP Headers with $\gamma_1:\gamma_2 = 1:3$}\\
    %     Ground-truth & &  & &  &  &  &  &  &  & \\
    %     GSExtract &8023/18911/8029  &6314/15617/6323 &5013/12918/5016  &3977/10562/3980  &3323/8851/3324  &2618/7040/2620  &2074/5427/2076  &1750/4509/1753  &1437/3626/1438  &1187/2956/1189\\
    %     CLExtract &10059/16734/8587 &9981/16758/8587  &9911/16755/8587 &9822/16776/8587  &9695/16697/8587 &9608/16723/8588  &9468/16704/8587  &9354/16705/8587  &9233/16672/8589  &9149/16724/8589 \\
    % \midrule
    %     & \multicolumn{9}{c}{\# of BB / GSE / IP Headers with $\gamma_1:\gamma_2 = 3:1$}\\\\
    %     Ground-truth & &  & &  &  &  &  &  &  & \\
    %     GSExtract &6446/16942/6449 &3950/11087/3952 &2628/7214/2628  &1730/4598/1733 &1104/2826/1104  &820/1981/820  &600/1411/600   &435/990/435  &290/639/290  &245/516/245 \\
    %     CLExtract  & 9920/16682/8587 &9686/16615/8587  &9412/16511/8588 &9057/16448/8587  &8706/16336/8589 &8373/16170/8588  &7994/15995/8589  &7553/15998/8588  &7114/15697/8588  &6503/15707/8589 \\
        \midrule
        & \multicolumn{9}{c}{Flip($\gamma_1$) : Loss($\gamma_2$) = $1:1$} \\
        GSExtract &17905 &13639 &9849 &7231  &5070  &3838 &2738  &1997  &1535 &1114 \\
        CLExtract &16684 &16663  &16637  &16601 &16541  &16444  &16461  &16340  &16253 &16275\\
    \midrule
        & \multicolumn{9}{c}{Flip($\gamma_1$) : Loss($\gamma_2$) = $1:3$}\\
        GSExtract &18911  &15617 &12918  &10562  &8851  &7040  &5427 &4509 &3626 &2956 \\
        CLExtract &16734 &16758  &16755 &16776  &16697 &16723  &16704  &16705  &16672  &16724 \\
    \midrule
        & \multicolumn{9}{c}{Flip($\gamma_1$) : Loss($\gamma_2$) = $3:1$}\\
        GSExtract &16942 &11087 &7214  &4598 &2826  &1981  &1411  &990  &639  &516 \\
        CLExtract  & 16682 &16615  &16511 &16448  &16336 &16170  &15995  &15998  &15697  &15707 \\
       \bottomrule
     
    \end{tabular}
    
    }
    
    \parbox{0.72\textwidth}{
    \vspace{1ex}
    
    \caption{Comparison with GSExtract in terms of the number of identified GSE headers, with different corruption degrees ($\gamma_1+\gamma_2$) and different ratios of corruption types ($\gamma_1 : \gamma_2$). The total number of GSE headers is 24287.}
    \label{tab:comparison}
    }
    \vspace{-4ex}
\end{table*}

\begin{figure}[t]
    \centering
    \begin{subfigure}[b]{0.26\textwidth}
         \includegraphics[width=1.4in]{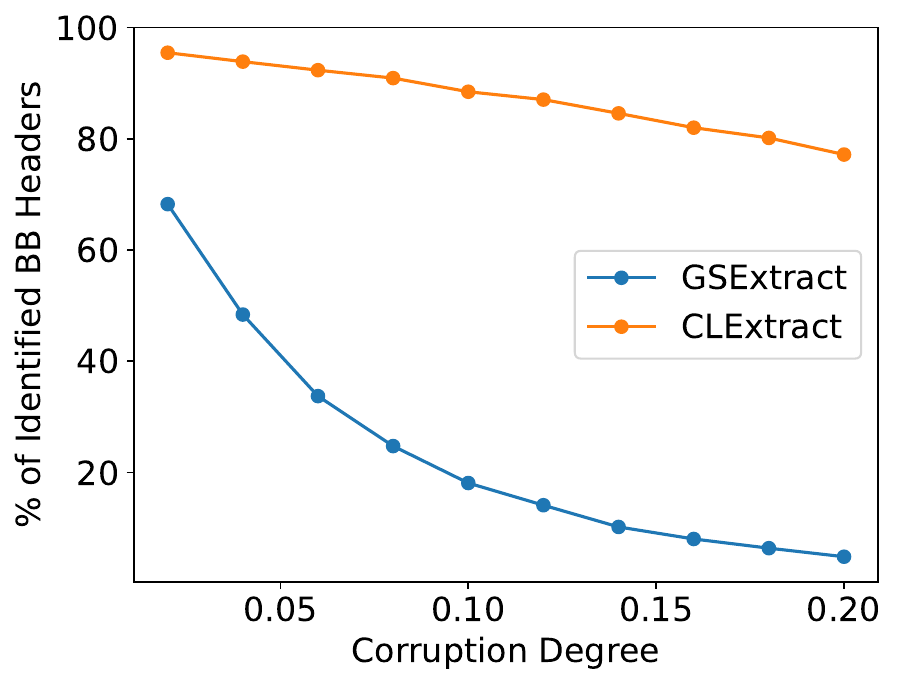}
         \caption{$\gamma_1$:$\gamma_2$ = $1:1$}
         \label{fig:ablation:gse:acc}
     \end{subfigure}
     \parbox{0.48\textwidth}{
    \caption{Corresponds to Table~\ref{tab:comparison}, comparison between \sys and GSExtract in BB header identification with different corruption degrees ($\gamma_1 + \gamma_2$) and fixed ratio ($\gamma_1 : \gamma_2$). More corresponding results are in Figure~\ref{fig:compare:bb} and \ref{fig:compare:ip}.}
    \label{fig:compare:bb:1:1}
    }
    \vspace{-5ex}
\end{figure}

\nip{Robustness.} 
We measure the robustness of \sys against different corruption degrees and ratios of corruption types.
Still, from Table~\ref{tab:effectiveness}, we can see that the four effectiveness metrics are not lowered too much when corruption becomes more severe.
Taking BB header as an example, its ACC, Precision, Recall, and F1 only drop 7\%, 2\%, 12\%, and 7\%, respectively, when the corruption degree increases from 2\% to 20\%. 
Similar trends are also presented in Figure~\ref{fig:exp1:bb}, \ref{fig:exp1:gse}, and \ref{fig:exp1:ip} that correspond to Table~\ref{tab:effectiveness}.

From Table~\ref{tab:comparison}, we can see that, in general, \sys performs well when the ratio of corruption types changes.
More specifically, \sys identifies 16734 GSE headers when the corruption degree is 2\% and flip($\gamma_1$) : loss($\gamma_2$) is 1:3. 
If we fix the corruption degree and switch the ratio to 3:1, the number is 16682 which is only 50 fewer.
When the corruption degree increases to 20\%, this gap becomes 1017 which takes over 4.2\% of the total number of GSE headers.
In comparison, GSExtract is significantly influenced by the ratio of corruption types. 
From Table~\ref{tab:comparison}, we can observe that the number of identified GSE headers drops 83.3\% (2956 to 516) if the ratio switches from 1:3 to 3:1.
It indicates that GSExtract is more likely to be dis-functioned by bit flip than bit loss.
This is because when a bit loss happens, there is still a 50\% chance that the value is correct.
However, when a bit flip happens, the value becomes completely wrong.
For GSExtract which heavily relies on critical information in headers, a wrong value can ruin the following decoding, leading to the missing of many headers. 
While for \sys, it learns header features as a whole and is less influenced by the information loss of bit flip.

In Figure~\ref{fig:compare:bb:1:1}, we present the comparison in BB header identification with fixed $\gamma_1 : \gamma_2$.
We can observe that when the corruption degree is 0.2, GSExtract almost malfunctions while \sys can identify around 80\% BB headers. Similar results are shown in BB header and IP header identification with different ratios (Figure~\ref{fig:compare:bb} and \ref{fig:compare:ip} in Appendix~\ref{appendix}).
As such, we conservatively conclude that \sys shows strong robustness and can recover satellite streams even if it is corrupted up to at least 20\%.

\nip{Data Augmentation.}
Recall that we employ data augmentation to improve the robustness of \sys, here, we evaluate to which extent data augmentation achieves this goal.

In Figure~\ref{fig:ablation:bb} (and Figure~\ref{fig:ablation:gse} \ref{fig:ablation:ip} in Appendix~\ref{appendix}), we present the effectiveness and robustness of identifying BB headers, GSE headers, and IP headers, with and without data augmentation.
We can clearly see that data augmentation significantly improves the effectiveness and robustness of \sys with the orange line lying over the green line. 
The only exception is the Precision metric. 
Actually, this is within expectation. 
Intuitively, data augmentation blurs the boundary between corrupted headers and non-header streams.
Therefore, the model with data augmentation will mistakenly identify non-header streams as corrupted headers, increasing \Code{FP} when calculating the Precision metric (\Code{TP/(TP+FP)}).
However, it does not mean that data augmentation makes \sys perform poorer.
On the one hand, the difference between Precision with and without data augmentation is less than 5\%. 
On the other hand, \sys aims to balance between Precision and Recall (\ie not missing \Code{TP} and \Code{FN} at the same time).
Therefore, in terms of F1 which is a comprehension of Precision and Recall, data augmentation improves the effectiveness and robustness.

\begin{figure}[t]
    \centering
    \begin{subfigure}[b]{0.26\textwidth}
         \includegraphics[width=1.4in]{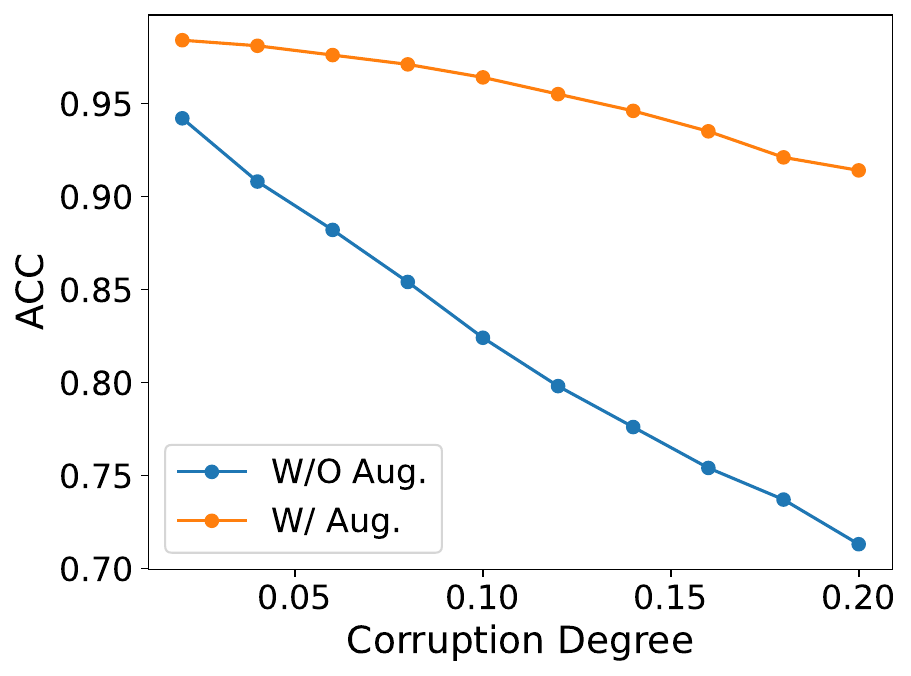}
         \subcaption{ACC}
         \label{fig:ablation:bb:acc}
     \end{subfigure}
    \hspace{-3em}
    \begin{subfigure}[b]{0.26\textwidth}
         \includegraphics[width=1.4in]{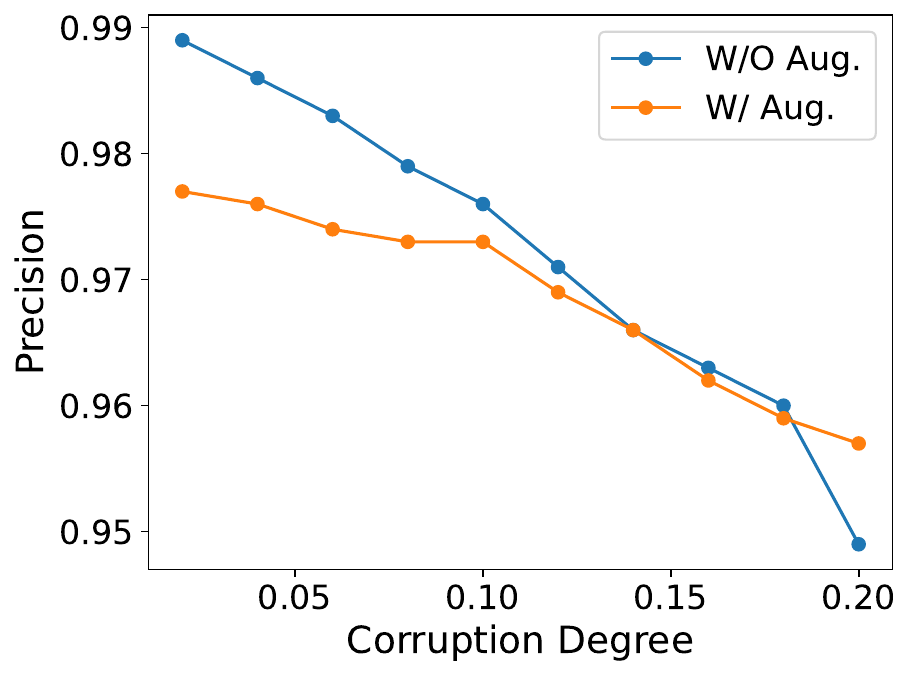}
         \subcaption{Precision}
         \label{fig:ablation:bb:pre}
     \end{subfigure}

    \begin{subfigure}[b]{0.26\textwidth}
         \includegraphics[width=1.4in]{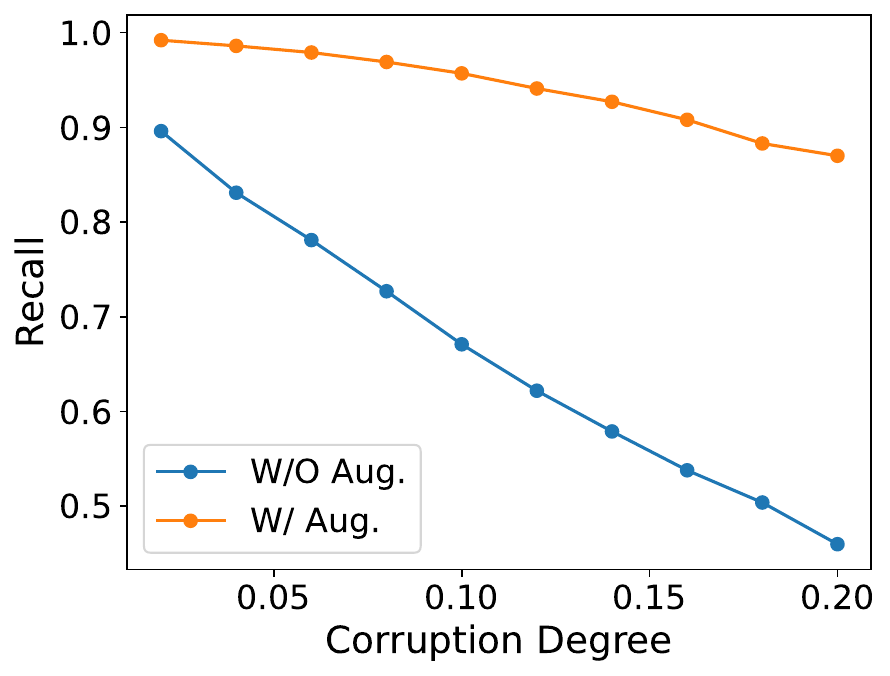}
         \subcaption{Recall}
         \label{fig:ablation:bb:rec}
     \end{subfigure}
    \hspace{-3em}
    \begin{subfigure}[b]{0.26\textwidth}
         \includegraphics[width=1.4in]{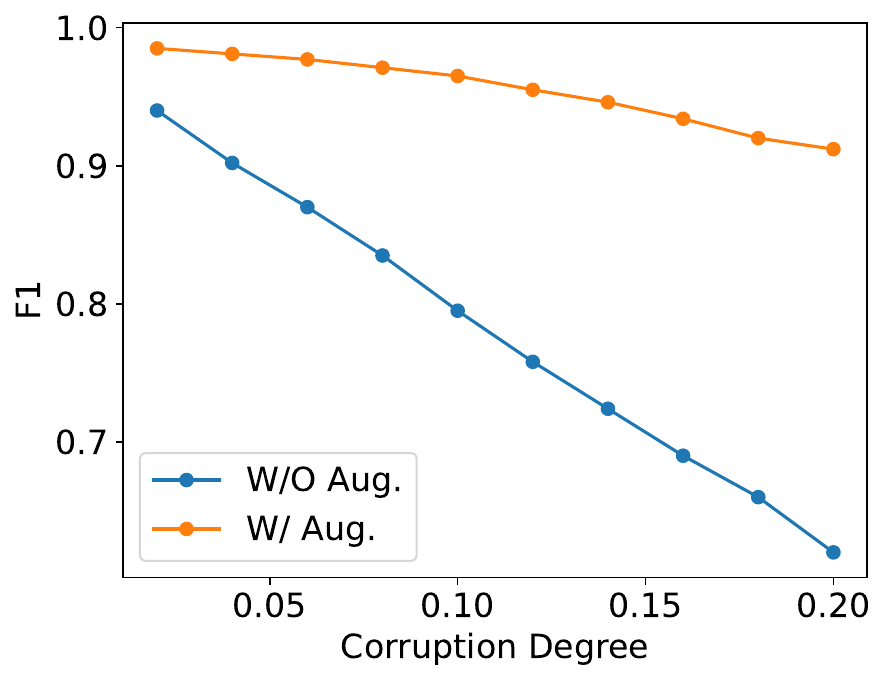}
         \subcaption{F1}
         \label{fig:ablation:bb:f1}
     \end{subfigure}
    \parbox{0.48\textwidth}{
    \caption{Effectiveness and robustness of BB header identification w/ data augmentation ($\gamma_1+\gamma_2$ =20\%)}
        \label{fig:ablation:bb}
    }
    \vspace{-3ex}
\end{figure}

\nip{Efficiency.} 
In our evaluation, we compare \sys with GSExtract in terms of efficiency. 
With all eavesdropped data, it takes GSExtract 10, 391.52 seconds to process and identify in total 10,471 BB headers, 24,287 GSE headers, and 10,479 IP headers - our training and testing set.
In other words, the identification rate is 0.1 BB headers, 0.23 GSE headers, and  0.1 IP headers per second. 

Using this dataset, it takes near 14.5 hours to train all models in \sys.
Applying our models, from the whole eavesdropped data, \sys spends 19,870.47 seconds processing and identifying in total 498,819 BB headers, 1,156,412 GSE headers, and 497,566 IP headers. 
That is to say, \sys can identify 25.10 BB headers, 58.20 GSE headers, and 25.04 IP headers per second. 
Admitted that the headers pinpointed by \sys can be false positives, from the statistics above, we can roughly estimate that \sys is hundreds of times faster than GSExtract.
In addition to more headers identified by \sys, another reason for the efficiency improvement is that \sys can be paralleled while GSExtract must perform decoding in a sequential fashion.
% \vspace{-1ex}
\section{Limitation and Future Work}
\label{sec:related}
This work has the following limitations. 
First, the effectiveness of GSE header identification is not as good as that of BB header identification and IP header identification. The reasons, as discussed in Section~\ref{subsec:eval-result}, are the variable length and the small size nature of GSE header. 
To address this problem, in the future, we plan to include bounding box regression and intersection over union techniques~\cite{redmon2016you} to further improve \sys.
Second, we evaluate the robustness of GSExtract and \sys up to 20\% corruption degree. 
We don't measure the performance when the corruption becomes more severe because 20\% is high enough to render the data payload meaningless. 
However, as future work, we plan to further raise the corruption degree to examine how \sys performs in an extreme case.
Third, to compare the efficiency of GSExtract and \sys, we count the numbers of identified headers in one second. These numbers are accurate for GSExtract but include false positives for \sys. 
We cannot eliminate false positives because we don't have the ground-truth of eavesdropped satellite streams. 
Though we can obtain a rough estimation from these numbers, to do compare more precisely, in the future, we plan to simulate the radio communication with the ground-truth. 
We will continue to open source our simulation platform to foster future research.
\section{Conclusion}
\label{sec:conclusion}
In this work, we design \sys to decode and recover highly corrupted satellite streams. \sys uses a contrastive learning technique with data augmentation to learn the features of packet headers at different protocol layers and identify them in a stream sequence.
By filtering them out, \sys extracts the innermost data payload that can be further analyzed by tools like Wireshark. 
Compared with the state-of-the-art GSExtract, \sys can successfully recover 71-99\% more eavesdropped data hundreds of times faster than GSExtract. 
Moreover, the effectiveness of \sys is not largely damaged when corruption becomes more severe.

% \newpage

% \end{thebibliography}

\bibliographystyle{IEEEtranS}
\bibliography{ref}
\clearpage
\newpage
\appendix
\label{appendix}

\begin{figure}[h]
    \centering
    \begin{subfigure}[b]{0.26\textwidth}
         \includegraphics[width=1.4in]{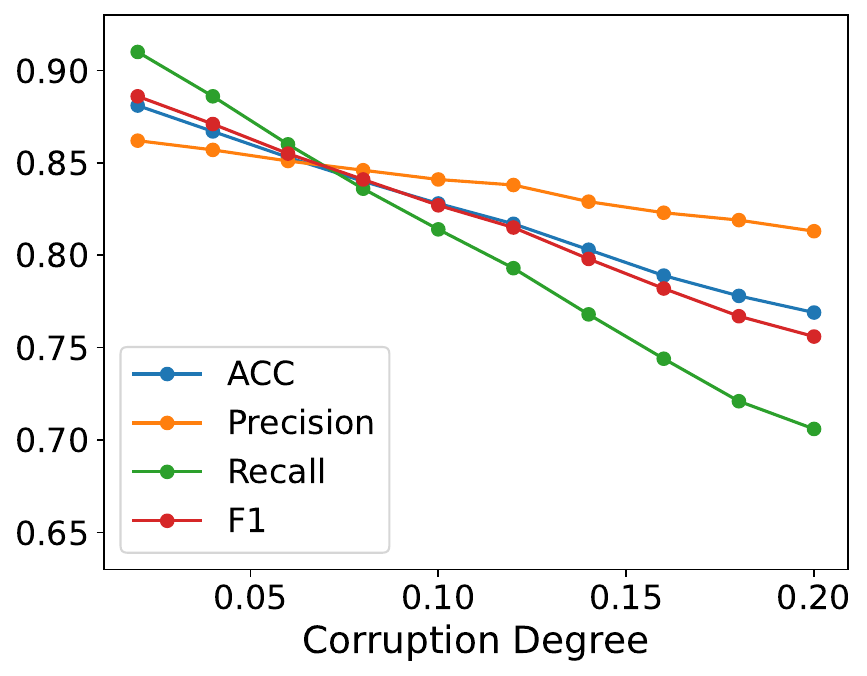}
         \caption{$\gamma_1$:$\gamma_2$ = $1:3$}
         \label{fig:exp1:gse:r25}
     \end{subfigure}
    \hspace{-3em}
    \begin{subfigure}[b]{0.26\textwidth}
         \includegraphics[width=1.4in]{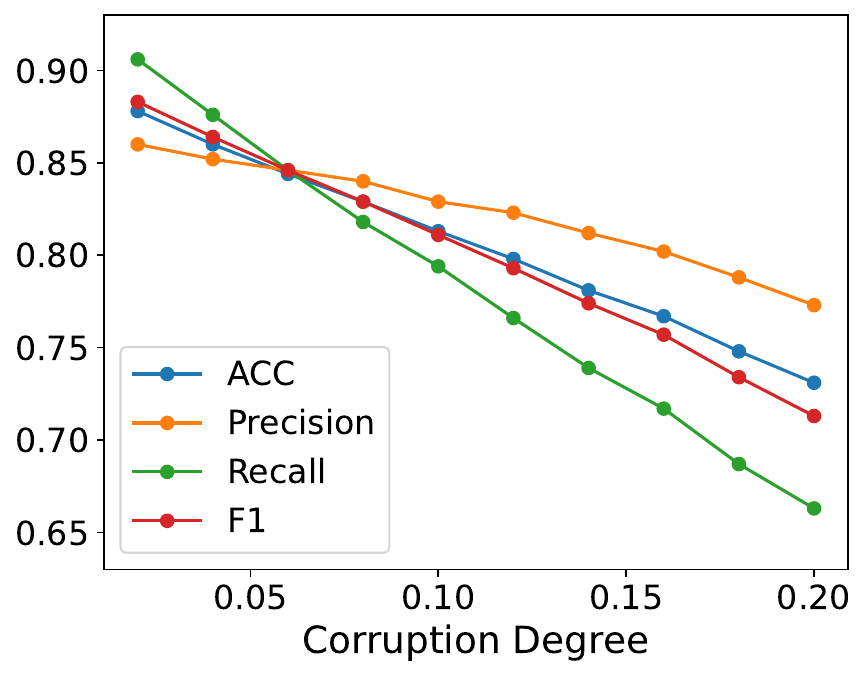}
         \caption{$\gamma_1$:$\gamma_2$ = $3:1$}
         \label{fig:exp1:gse:r75}
     \end{subfigure}
    \parbox{0.48\textwidth}{
    \caption{Corresponds to Table~\ref{tab:effectiveness}, effectiveness and robustness of GSE header identification with different corruption degrees ($\gamma_1 + \gamma_2$) and ratios ($\gamma_1 : \gamma_2$).}
    \label{fig:exp1:gse}
        }
\end{figure}

\begin{figure}[h]
    \centering
    \begin{subfigure}[b]{0.26\textwidth}
         \includegraphics[width=1.4in]{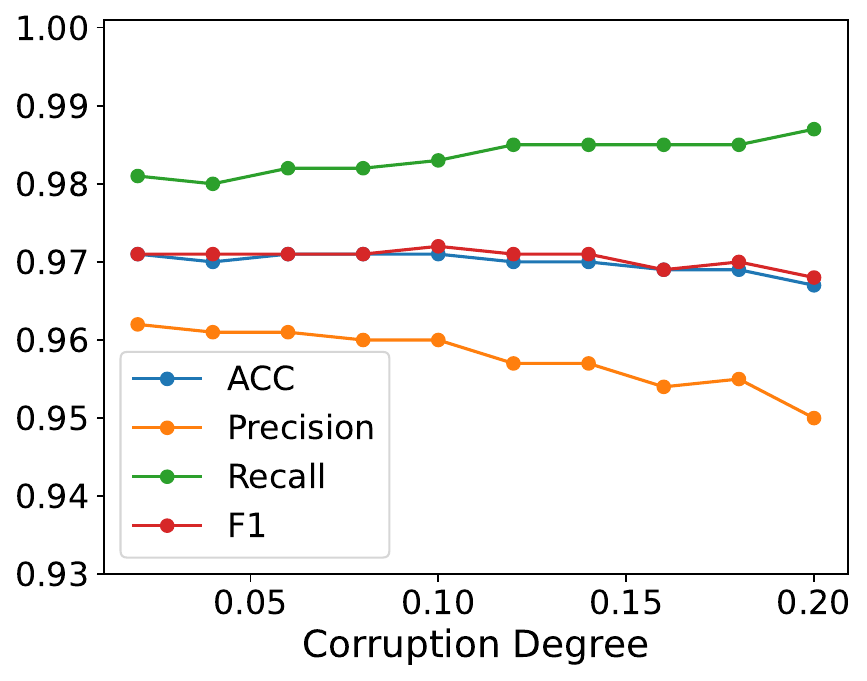}
         \caption{$\gamma_1$:$\gamma_2$ = $1:3$}
         \label{fig:exp1:ip:r25}
     \end{subfigure}
    \hspace{-3em}
    \begin{subfigure}[b]{0.26\textwidth}
         \includegraphics[width=1.4in]{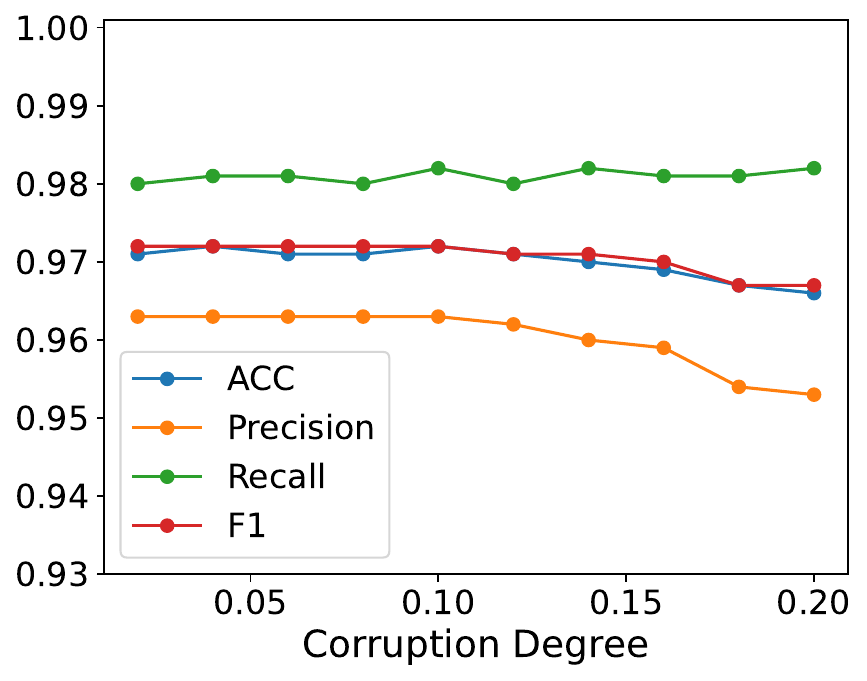}
         \caption{$\gamma_1$:$\gamma_2$ = $3:1$}
         \label{fig:exp1:ip:r75}
     \end{subfigure}
    \parbox{0.48\textwidth}{
    \caption{Corresponds to Table~\ref{tab:effectiveness}, effectiveness and robustness of IP header identification with different corruption degrees ($\gamma_1 + \gamma_2$) and ratios ($\gamma_1 : \gamma_2$)}
    \label{fig:exp1:ip}
        }
\end{figure}

\begin{figure}[h]
    \centering
    \begin{subfigure}[b]{0.26\textwidth}
         \includegraphics[width=1.4in]{figs/exp/headerNumRatio/1_1_bbheader.pdf}
         \caption{$\gamma_1$:$\gamma_2$ = $1:1$}
         \label{fig:ablation:gse:acc}
     \end{subfigure}
    \hspace{-3em}
    \begin{subfigure}[b]{0.26\textwidth}
         \includegraphics[width=1.4in]{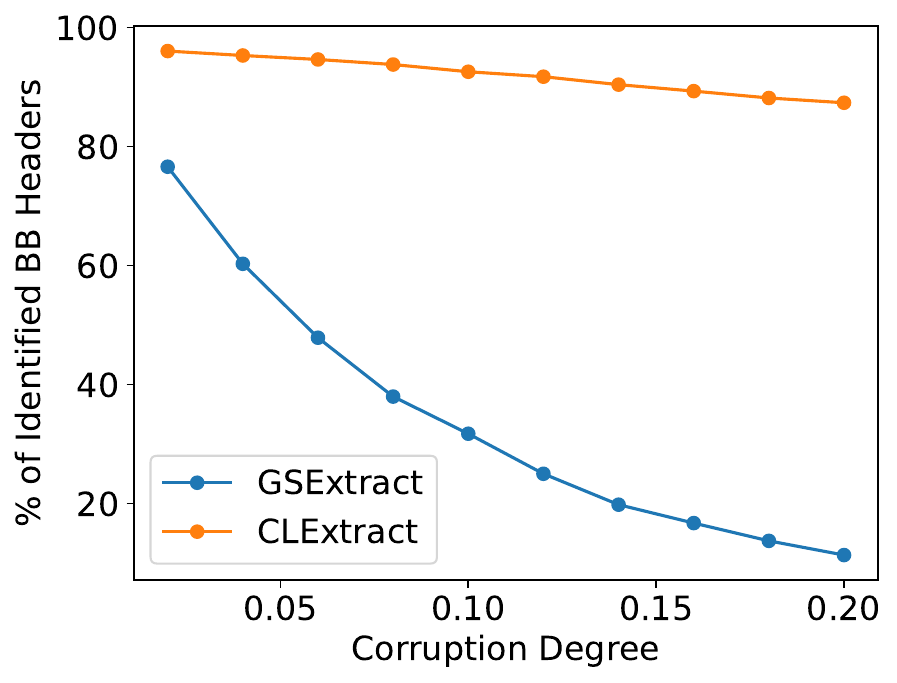}
         \caption{$\gamma_1$:$\gamma_2$ = $1:3$}
         \label{fig:ablation:gse:pre}
     \end{subfigure}

    \begin{subfigure}[b]{0.26\textwidth}
         \includegraphics[width=1.4in]{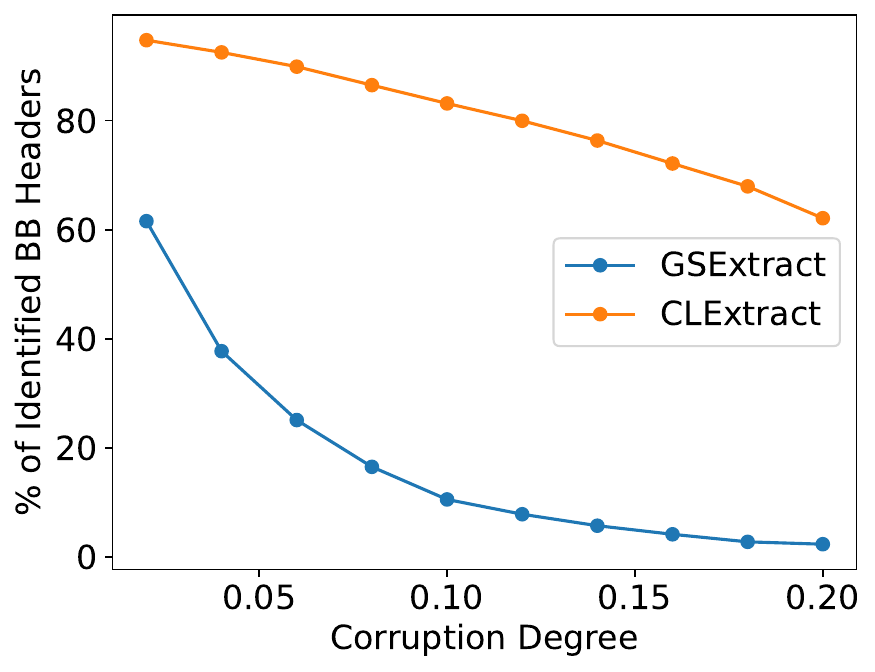}
         \caption{$\gamma_1$:$\gamma_2$ = $3:1$}
         \label{fig:ablation:gse:rec}
     \end{subfigure}
     \parbox{0.48\textwidth}{
    \caption{Corresponds to Table~\ref{tab:comparison}, comparison between \sys and GSExtract in BB header identification with different corruption degrees ($\gamma_1 + \gamma_2$) and ratios ($\gamma_1 : \gamma_2$).}
    \label{fig:compare:bb}
    }
\end{figure}

\begin{figure}[h]
    \centering
    \begin{subfigure}[b]{0.26\textwidth}
         \includegraphics[width=1.4in]{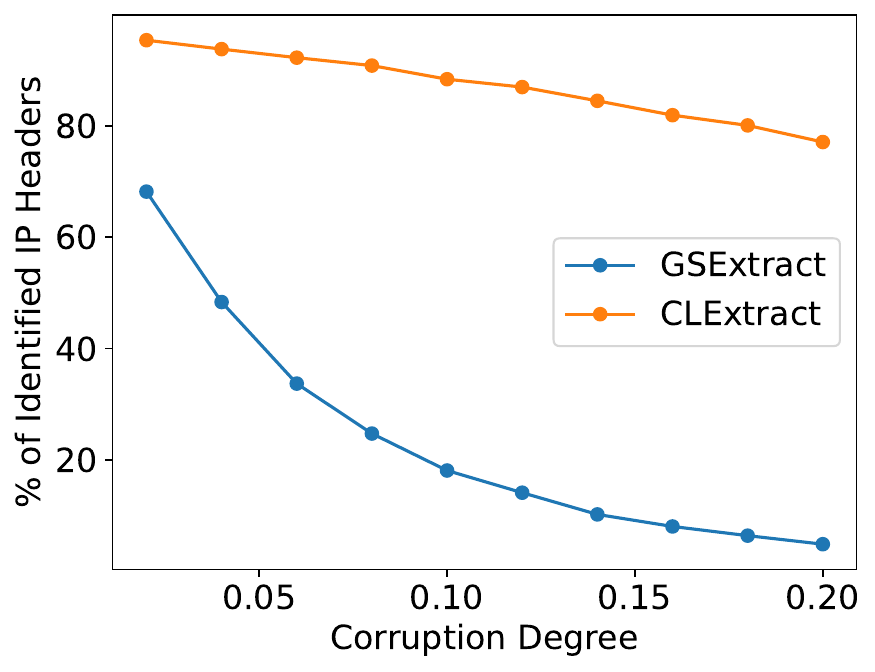}
         \caption{$\gamma_1$:$\gamma_2$ = $1:1$}
         \label{fig:ablation:gse:f1}
     \end{subfigure}
     \hspace{-3em}
     \begin{subfigure}[b]{0.26\textwidth}
         \includegraphics[width=1.4in]{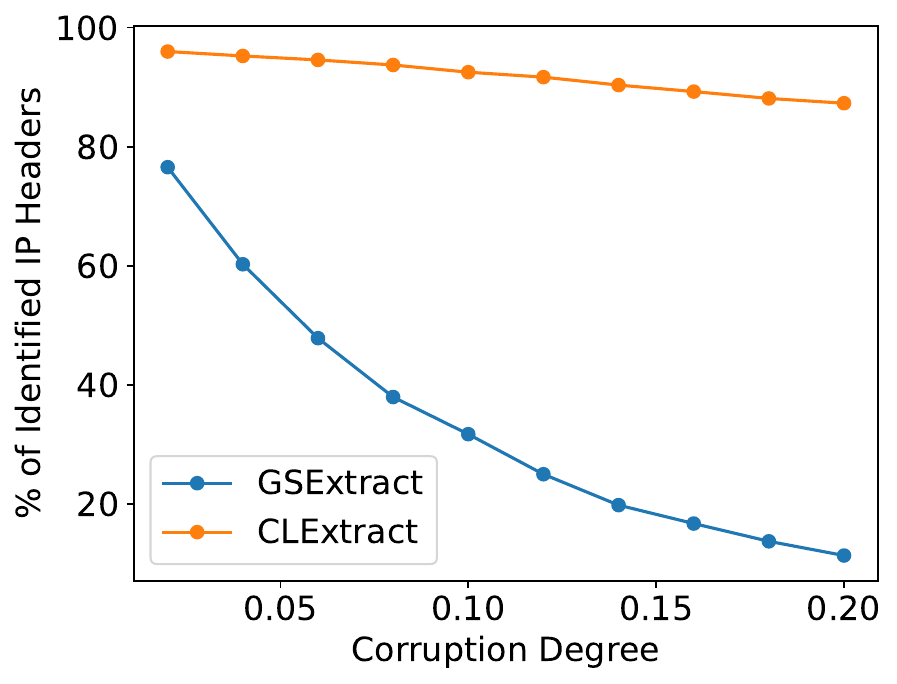}
         \caption{$\gamma_1$:$\gamma_2$ = $1:3$}
         \label{fig:ablation:gse:f1}
     \end{subfigure}
    \begin{subfigure}[b]{0.26\textwidth}
         \includegraphics[width=1.4in]{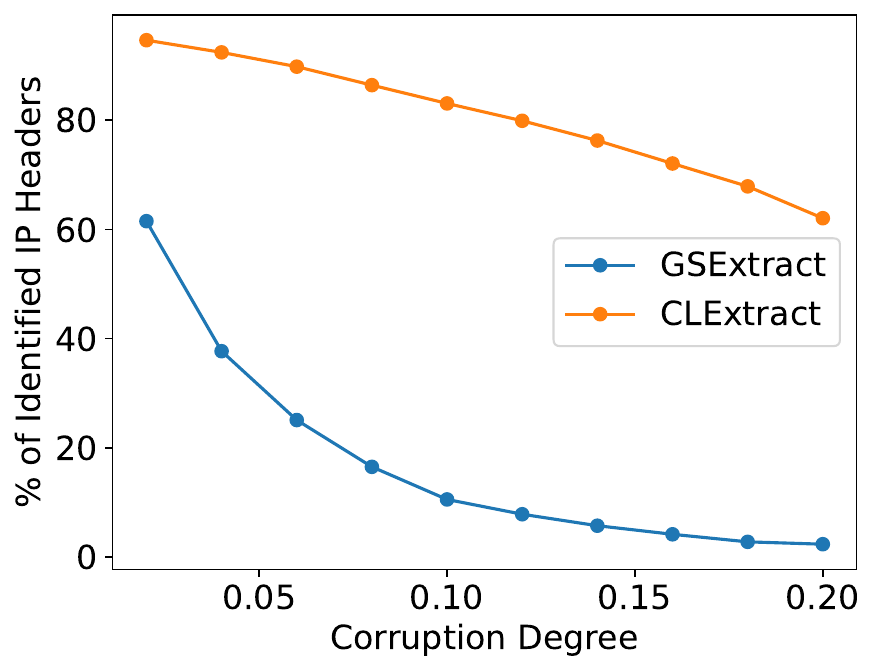}
         \caption{$\gamma_1$:$\gamma_2$ = $3:1$}
         \label{fig:ablation:gse:f1}
     \end{subfigure}
    \parbox{0.48\textwidth}{
    \caption{Corresponds to Table~\ref{tab:comparison}, comparison between \sys and GSExtract in IP header identification with different corruption degrees ($\gamma_1 + \gamma_2$) and ratios ($\gamma_1 : \gamma_2$).}
    \label{fig:compare:ip}
    }
\end{figure}

\begin{figure}[h]
    \centering
    \begin{subfigure}[b]{0.26\textwidth}
         \includegraphics[width=1.4in]{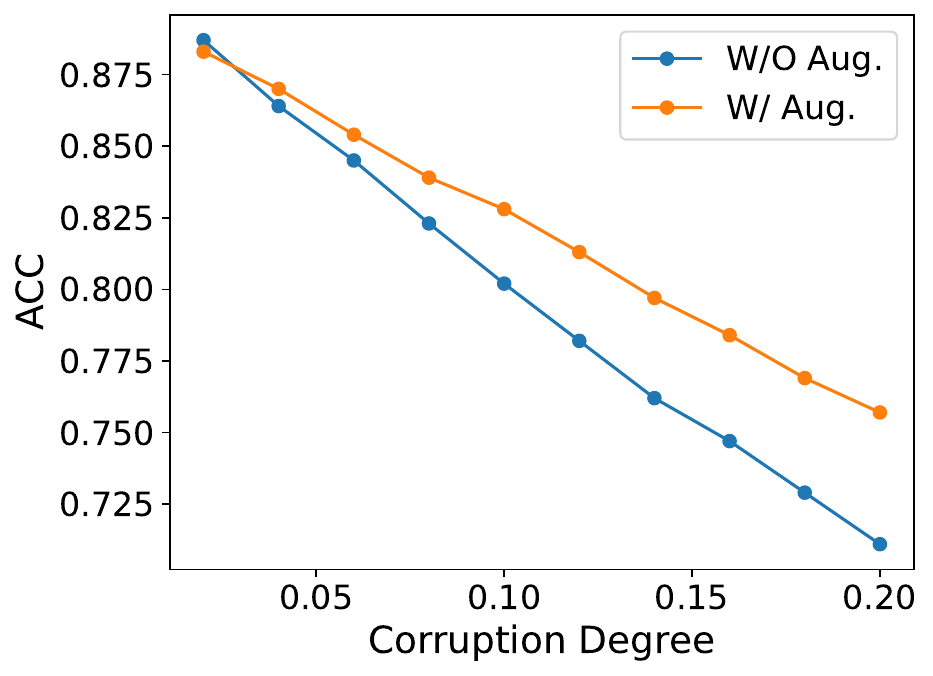}
         \caption{ACC}
         \label{fig:ablation:gse:acc}
     \end{subfigure}
    \hspace{-3em}
    \begin{subfigure}[b]{0.26\textwidth}
         \includegraphics[width=1.4in]{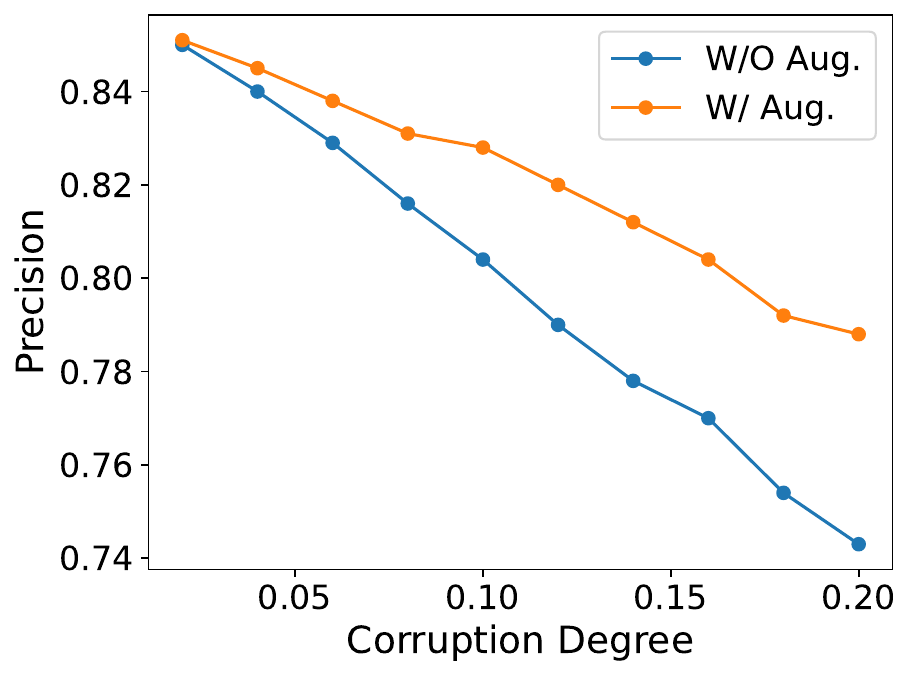}
         \caption{Precision}
         \label{fig:ablation:gse:pre}
     \end{subfigure}
    \begin{subfigure}[b]{0.26\textwidth}
         \includegraphics[width=1.4in]{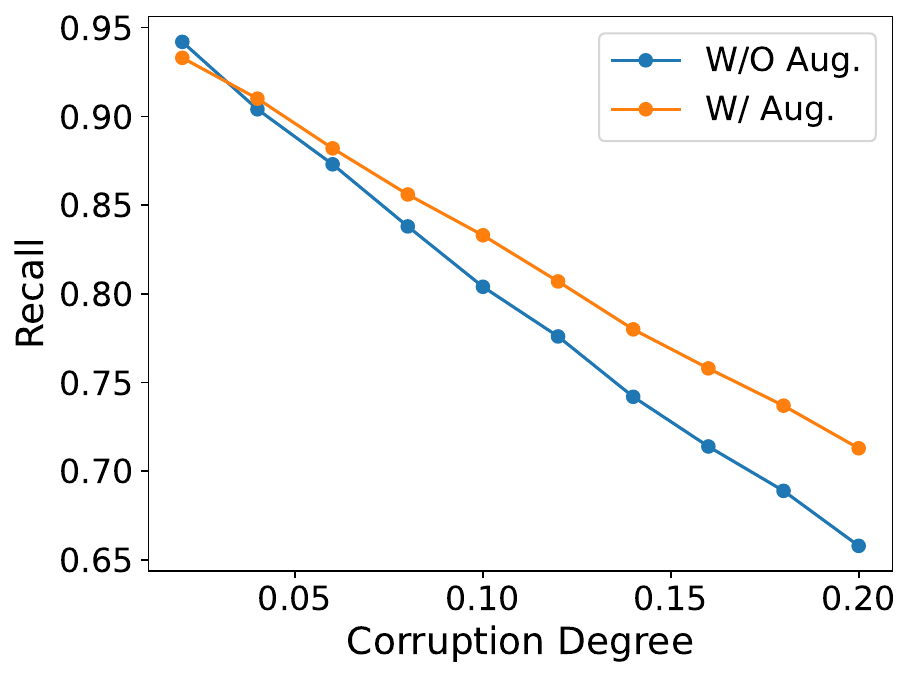}
         \caption{Recall}
         \label{fig:ablation:gse:rec}
     \end{subfigure}
    \hspace{-3em}
    \begin{subfigure}[b]{0.26\textwidth}
         \includegraphics[width=1.4in]{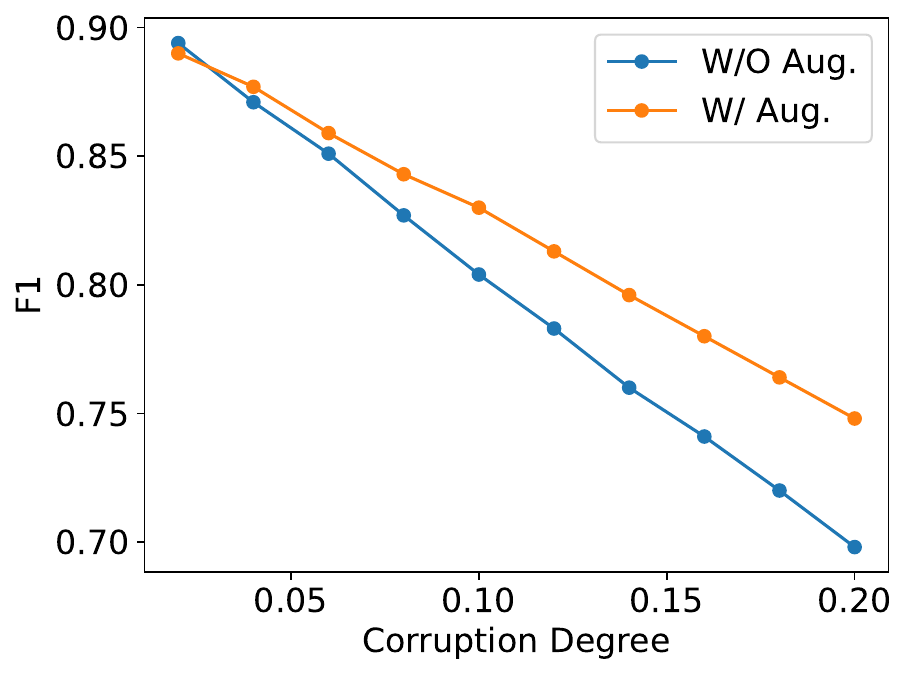}
         \caption{F1}
         \label{fig:ablation:gse:f1}
     \end{subfigure}
    \parbox{0.48\textwidth}{
    \caption{Corresponds to Figure~\ref{fig:ablation:bb}, effectiveness and robustness of GSE header identification w/ data augmentation ($\gamma_1+\gamma_2$ =20\%)}
    \label{fig:ablation:gse}
    }
\end{figure}

\begin{figure}[h]
    \centering
    \begin{subfigure}[b]{0.26\textwidth}
        \includegraphics[width=1.4in]{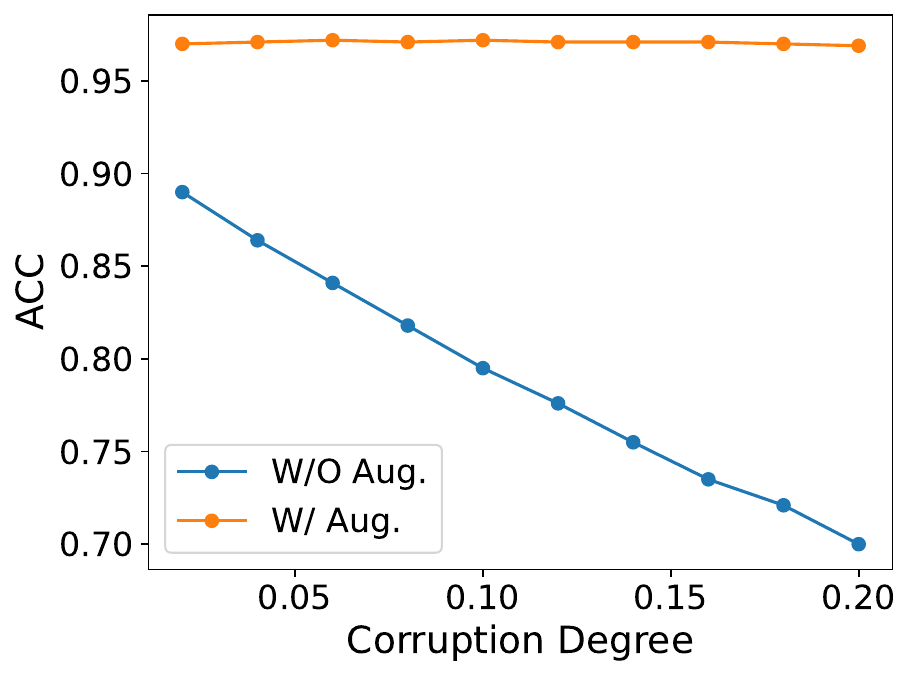}
         \caption{ACC}
         \label{fig:ablation:ip:acc}
     \end{subfigure}
    \hspace{-3em}
    \begin{subfigure}[b]{0.26\textwidth}
         \includegraphics[width=1.4in]{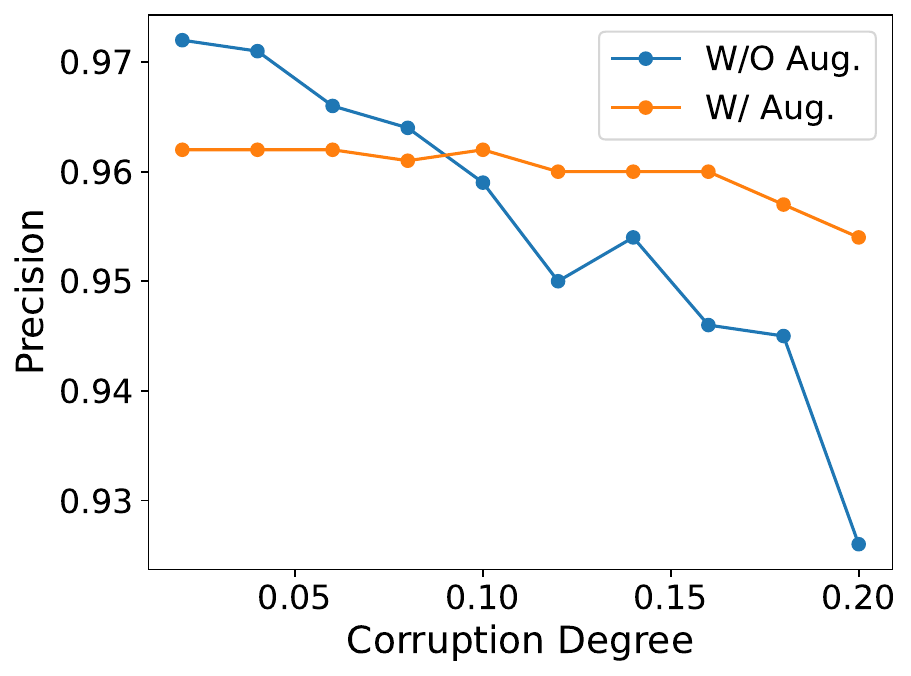}
         \caption{Precision}
         \label{fig:ablation:ip:pre}
     \end{subfigure}

    \begin{subfigure}[b]{0.26\textwidth}
         \includegraphics[width=1.4in]{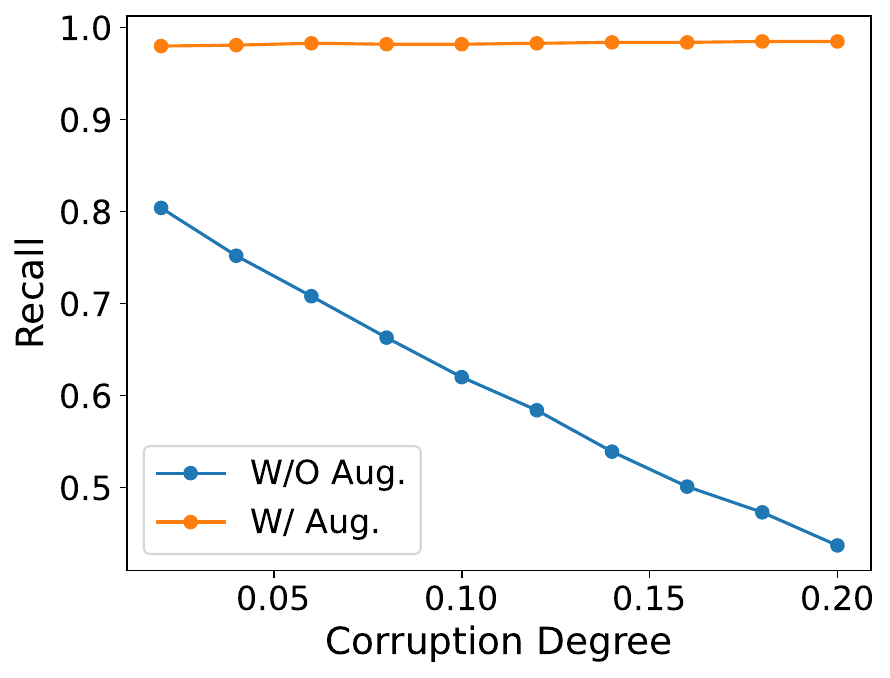}
         \caption{Recall}
         \label{fig:ablation:ip:rec}
     \end{subfigure}
    \hspace{-3em}
    \begin{subfigure}[b]{0.26\textwidth}
         \includegraphics[width=1.4in]{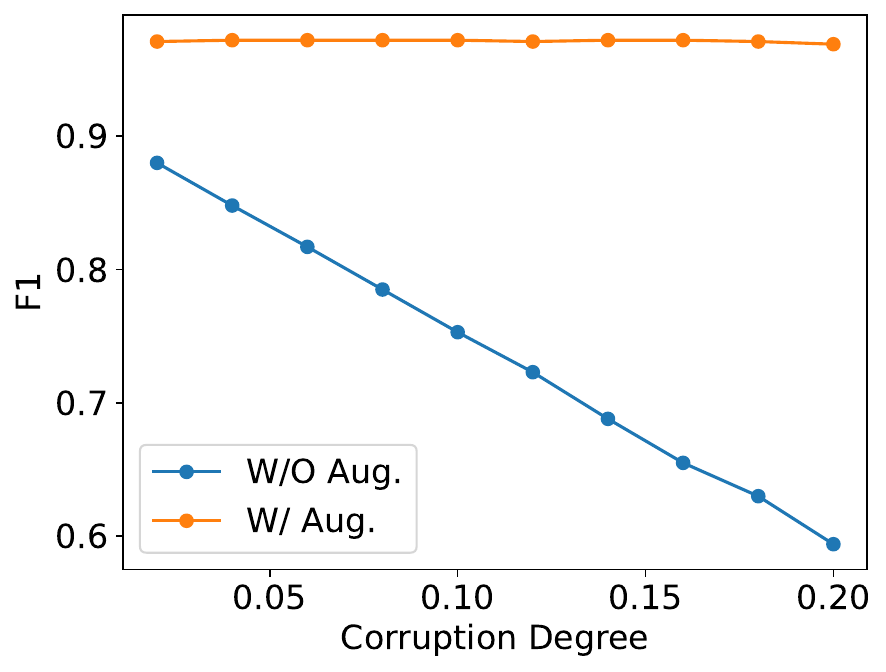}
         \caption{F1}
         \label{fig:ablation:ip:f1}
     \end{subfigure}
    \parbox{0.48\textwidth}{
    \caption{Corresponds to Figure~\ref{fig:ablation:bb}, effectiveness and robustness of IP header identification w/ data augmentation ($\gamma_1+\gamma_2$ =20\%)}
        \label{fig:ablation:ip}
    }
\end{figure}

Due to the space limit, we moved some evaluation results to the Appendix. 
For effectiveness and robustness evaluated over testing data, refer to Figure~\ref{fig:exp1:gse} and \ref{fig:exp1:ip}.
For comparison between \sys and GSExtract, refer to  Figure~\ref{fig:compare:bb} and \ref{fig:compare:ip}.
For data augmentation, refer to  Figure~\ref{fig:ablation:gse} and \ref{fig:ablation:ip}.

\newpage

\newpage

\end{document}